\documentclass[journal,transmag]{IEEEtran}
\usepackage{dsfont}
\usepackage{algorithmic}
\usepackage{algorithm}
\usepackage{bm}
\usepackage{amsmath}
\usepackage{subfigure}
\usepackage{multirow}
\usepackage{caption}
\usepackage{amsmath}
\usepackage{booktabs}
\usepackage{subfigure}
\usepackage{caption}
\usepackage{tabularx}
\usepackage{graphicx}
\usepackage{stfloats}

\usepackage{array}
\newcommand{\PreserveBackslash}[1]{\let\temp=\\#1\let\\=\temp}
\newcolumntype{C}[1]{>{\PreserveBackslash\centering}p{#1}}
\newcolumntype{R}[1]{>{\PreserveBackslash\raggedleft}p{#1}}
\newcolumntype{L}[1]{>{\PreserveBackslash\raggedright}p{#1}}

\newlength{\halfpagewidth}
    \divide\halfpagewidth by 2

\ifCLASSINFOpdf
\else
\fi
\begin{document}
\onecolumn
%
\title{Sparse Representation Based Efficient Radiation Symmetry Analysis Method for Cylindrical Model of Inertial Confinement Fusion}


\author{\IEEEauthorblockN{Yanfeng Zhang,
}
School of Electronic and Information Engineering, Harbin Institute of Technology, Shenzhen, P.R. China}

\markboth{Journal of \LaTeX\ Class Files,~Vol.~14, No.~8, August~2015}%
{Shell \MakeLowercase{\textit{et al.}}: Bare Demo of IEEEtran.cls for IEEE Transactions on Magnetics Journals}
%



\IEEEtitleabstractindextext{%
\begin{abstract}
Radiation symmetry evaluation is critical to the laser driven Inertial Confinement Fusion (ICF), which is usually done by solving a view-factor equation model. The model is nonlinear, and the number of equations can be very large when the size of discrete mesh element is very small to achieve a prescribed accuracy, which may lead to an intensive equation solving process. In this paper, an efficient radiation symmetry analysis approach based on sparse representation is presented, in which, 1) the Spherical harmonics, annular Zernike polynomials and Legendre-Fourier polynomials are employed to sparsely represent the radiation flux on the capsule and cylindrical cavity, and the nonlinear energy equilibrium equations are transformed into the equations with sparse coefficients, which means there are many redundant equations, 2) only a few equations are selected to recover such sparse coefficients with Latin hypercube sampling, 3) a Conjugate Gradient Subspace Thresholding Pursuit (CGSTP) algorithm is then given to rapidly obtain such sparse coefficients equation with as few iterations as possible. Finally, the proposed method is validated with two experiment targets for Shenguang II and Shenguang III laser facility in China. The results show that only one tenth of computation time is required to solve one tenth of equations to achieve the radiation flux with comparable accuracy. Further more, the solution is much more efficient as the size of discrete mesh element decreases, in which, only 1.2\% computation time is required to obtain the accurate result.
\end{abstract}

\begin{IEEEkeywords}
radiation symmetry, inertial confinement fusion, sparse representation, compressed sensing.
\end{IEEEkeywords}}

\maketitle

\IEEEdisplaynontitleabstractindextext

%
\IEEEpeerreviewmaketitle

\section{Introduction}
%
%
%
%
\IEEEPARstart{C}{ontrollable} nuclear fusion is potential to solve the energy crisis in the future, and laser driven Inertial Confinement Fusion (ICF) is supposed to be one of the most promising ways \cite{rj1}. To achieve this goal, the driven asymmetry on the centrally located capsule needs to be evaluated and remains no more than 1\% during the fusion process. Due to limited laser beams for NIF laser facility, i.e., 2 laser entrance holes and a few discrete diagnose holes, the radiation asymmetry may be much larger than such prescribed threshold \cite{rj2}. Therefore, we need to efficiently compute the driven flux reached the capsule to evaluate its driven asymmetry.
	In the ICF, the radiation flux on the fuel capsule is related to the laser-plasma interactions (LPI) and transport of X-rays from the cavity wall to the capsule, which involves the solving of complex kinetic and hydrodynamic equations implemented in the codes such as LASNEX \cite{rj3}. In practice, simple mathematical models such as view-factor codes are usually employed to compute the radiation flux distributed on the capsule, especially for the preliminary design and optimization of thermonuclear target structure and shape \cite{rj3,rj4}. As described in \cite{rj1}, the radiation flux is usually obtained by solving a non-linear view-factor based equation model, which can be solved by utilizing Newton-Rapshon \cite{rj5}, Jacobi iteration \cite{rj6}, Cholesky decomposition \cite{rj7}, or Preconditioned Conjugate Gradient method \cite{rj8}. However, in order to improve the evaluation accuracy, the size of discrete element is usually very small, the number of elements or equations will increase significantly, which may lead to much time required to solve large scale nonlinear equations. The computation time may be unacceptable for researchers. Therefore, a new efficient computation approach is essentially required to significantly improve the efficiency of radiation symmetry evaluation and optimization.

	Compressed sensing (CS), proposed by Donoho and Candes \cite{rj9,rj10}, is a new method to reconstruct signals from significantly fewer measurements than that required by traditional methods, which has attracted considerable attention and achieves successful applications such as Medical imaging (MI) \cite{rj11}, Analog-digital Conversion \cite{rj12}, Computational Biology \cite{rj13}, and Computer Graphics \cite{rj14}. In the field of radiation symmetry evaluation of ICF, compressed analysis approach has been applied to efficiently obtain the radiation flux distribution on the capsule \cite{rj15}. However, the radiation flux computation model is simplified, and the accuracy of the radiation asymmetry evaluation is limited. The non-linear Time Dependent Energy Balance Model (TDEBM) presented in \cite{rj16} is often used to compute the radiation flux. Therefore, compressive sensing approach for non-linear equation solving approach, such as Iterative Hard Thresholding (IHT) \cite{rj17}, can be applied. However, the step size is often taken as a constant, which may lead to more times of iteration convergence. Normalized Iterative Hard Thresholding (NIHT) \cite{rj18} is further proposed, in which, fixed step length is replaced with a descending factor to accelerate the convergence rate. Nevertheless, the gradient based search direction in NIHT, may lead to the convergence is very slow when it approaches the minimum. Conjugate Gradient Iterative Hard Thresholding (CGIHT) is proposed in \cite{rj19}, to replace the search direction with the conjugate gradient direction, which can benefit accelerate the convergence. However, the search direction of CGIHT may not be conjugate, which may lead to more times of iterations being required. As one of the iterative greedy algorithms, Subspace Pursuit (SP) algorithm has attracted much attention for its backtracking idea in solving compressed sensing reconstruction problems, which requires only a small number of iterations \cite{rj20}. However, SP algorithm can not be directly used to solve nonlinear problems. In this paper, we absorb the advantages of CGIHT algorithm and SP algorithm, and propose a Conjugate Gradient Subspace Thresholding Pursuit (CGSTP) algorithm to largely reduce the TDEBM and efficiently obtain the radiation flux. The core ideas include:
\begin{enumerate}[leftmargin=8mm,labelsep=4mm]
\item[(1)] The spherical harmonics, annular Zernike polynomials and Legendre-Fourier polynomials are employed to sparsely represent the radiation flux on the capsule and spherical cavity, which enable only a few equations are required to recover the radiation flux with high accuracy, and significantly reduce the radiation computation model.

\item[(2)] Over the three sets of polynomials domain, such reduced nonlinear equation model with sparse coefficients is formulated to enable compressed sensing algorithms, such as CGIHT, be used to efficiently obtain the radiation flux. 

\item[(3)] Conjugate Gradient Subspace Thresholding Pursuit algorithm is then presented to enable the adjacent search direction be conjugate, which facilitate the convergence of iteration, rapidly obtain the solution, and then efficiently evaluate the radiation symmetry for ICF experiments design.
\end{enumerate}

The outline of this paper as follows. Section 2 introduces the nonlinear view-factor based equation model in a cylindrical cavity model. Radiation sparse representation model constructed with three orthogonal bases is described in section 3. Then section 4 gives a description of the proposed reconstruction algorithm, i.e. Conjugate Gradient Subspace Thresholding Pursuit (CGSTP) algorithm. At last, section 6 concludes following experiments validation in section 5.

%
\section{Sparse representation of radiation flux}
\begin{figure}
\centering
{\includegraphics[width=8 cm]{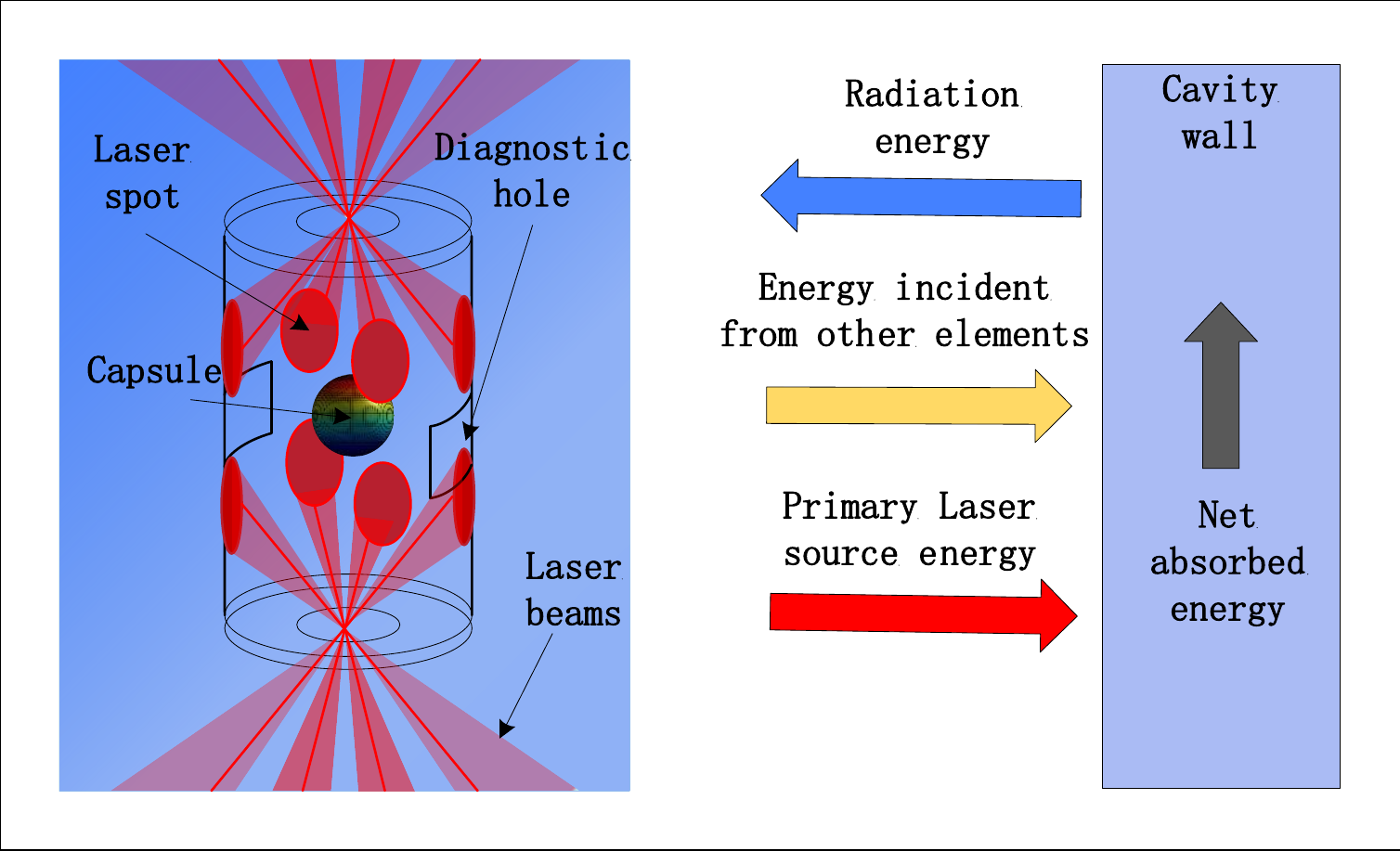}}
\caption{Scheme of the cylinder-to-sphere model and the energy exchange process for an element}\label{fig1}
\end{figure}

\subsection{Radiation flux computation model based on view-factor}

As shown in Figure \ref{fig1} , eight laser beams are injected into the cylindrical cavity through two entrance holes. By interacting with the high-Z material on the inner wall of the cavity, the laser incident on the cavity is converted into X-ray, the X-ray irradiates evenly and drives the centrally located capsule implosion. With generated discrete mesh elements, the radiation flux on the wall of the cavity and surface of the capsule can be computed by using the view-factor function described in \cite{rj15}, i.e. the TDEBM. For a discrete mesh element, according to the energy conservation law, the energy received from other elements equals the energy emits to other elements. It can be formulated as:

\begin{equation}
{\lambda _i}({E_i} + \sum\limits_{j = 1}^N {{F_{ji}}{\zeta _j}} {B_j}) = {B_i},
\label{eq.1}
\end{equation}
where $N$ is the total number of mesh elements, ${E_i}$ is the radiation flux comes directly from primary source converted from laser on the $i$-th mesh element which have a centroid ${{\bm{p}}_i}$ and normal ${{\bm{n}}_i}$, can be computed as ${E_i} = \sum\nolimits_{i = 0}^n {S_j^0} {F_{ji}}{\zeta _j}$, here, $S_j^0$ is the radiation flux converted directly from laser beams of the $j$-th element. ${\zeta _j}$ is the area of the $j$-th mesh element with a centroid ${{\bm{p}}_j}$ and normal ${{\bm{n}}_j}$. ${F_{ji}}$ is the view-factor between the $j$-th mesh element and the $i$-th mesh element, which can be computed by

\begin{equation}
{F_{ji}} = {{{\frac{1}{\pi }\left[ {{{\bm{n}}_i} \cdot \left( {{{\bm{p}}_j} - {{\bm{p}}_i}} \right)} \right] \cdot \left[ {{{\bm{n}}_j} \cdot \left( {{{\bm{p}}_i} - {{\bm{p}}_j}} \right)} \right]} \mathord{\left/
 {\vphantom {{\frac{1}{\pi }\left[ {{{\bf{n}}_i} \cdot \left( {{{\bm{p}}_j} - {{\bm{p}}_i}} \right)} \right] \cdot \left[ {{{\bm{n}}_j} \cdot \left( {{{\bm{p}}_i} - {{\bm{p}}_j}} \right)} \right]} {\left\| {{{\bm{p}}_i} - {{\bm{p}}_j}} \right\|}}} \right.
 \kern-\nulldelimiterspace} {\left\| {{{\bm{p}}_i} - {{\bm{p}}_j}} \right\|}}^4},
\label{eq.2}
\end{equation}
and ${\lambda _i}$ is the albedo of the $i$-th mesh element, which is relating to the wall material, radiation time $t$ and the emit energy ${B_i}$, which can be given as

\begin{equation}
{\lambda _i} = \frac{1}{{1 + {\upsilon ^{ - {\beta ^{ - 1}}}}B_i^{{\beta ^{ - 1}} - 1}{t^{ - \alpha {\beta ^{ - 1}}}}}},
\label{eq.3}
\end{equation}
where $\upsilon$ , $\alpha$ and $\beta$ are the material parameters in which $\upsilon  = 4.87$, $\alpha  = 8/13$, $\beta  = 16/13$. ${B_i}$ and ${B_j}$ are the radiation flux of the $i$-th mesh element and the $j$-th mesh element respectively. Let ${\bm{E}} = {\left[ {{E_1},{E_2}, \cdot  \cdot  \cdot ,{E_n}} \right]^{\rm T}}$ , ${\bm{S}} = {\left[ {S_1^0,S_2^0, \cdot  \cdot  \cdot ,S_N^0} \right]^{\rm T}}$, ${\bm{F}} = ({F_{ji}})$, ${\bm{\zeta }} = {\left[ {{\zeta _1},{\zeta _2}, \cdot  \cdot  \cdot ,{\zeta _N}} \right]^{\rm T}}$, ${\bm{B}} = {\left[ {{B_1},{B_2}, \cdot  \cdot  \cdot ,{B_N}} \right]^{\rm T}}$, $C = {\upsilon ^{ - {\beta ^{ - 1}}}}{t^{ - \alpha {\beta ^{ - 1}}}}$, where ${\rm T}$ indicates the matrix transpose. Let ${\bm{V}} = ({F_{ji}}{\zeta _j})$  be an intermediate variable, then the matrix form of ${E_i} = \sum\nolimits_{i = 0}^n {S_j^0} {F_{ji}}{\zeta _j}$ can be written as ${\bm{E}} = {\bm{VS}}$ and Equation \eqref{eq.3} can be written in a compact form as

\begin{equation}
\label{eq.4}
({\bm{I}} - {\bm{V}}){\bm{B}} + C{\bm{B}}{{^\circ}^{({1 \mathord{\left/
 {\vphantom {1 \beta }} \right.
 \kern-\nulldelimiterspace} \beta )}}} = {\bm{E}}
\end{equation}
where $\bm{I}$ denotes an unit matrix. In Equation \eqref{eq.4}, $\bm{V}$ and $\bm{E}$ are known, $\bm{B}$ is the unique unknown variable. $^\circ $ denote the hadamard power which means for any vector $\bm{{\cal V}}$ and constant $a$, $\bm{{\cal V}}{{^\circ} ^a} = {\left[ {{\cal V}_1^a,{\cal V}_2^a, \cdot  \cdot  \cdot ,{\cal V}_n^a} \right]^{\rm T}}$ holds.
Since $\beta$ do not equal one, Equation \eqref{eq.4} is nonlinear. Such nonlinear equations can be solved by utilizing some conventional iterative methods such as Newton-Raphson method. However, these methods is time consuming even unusable when the number of the discrete elements reaches ${10^7}$ or more. 


\subsection{Compressed Sensing Method}
Compressed Sensing (CS) also called Compressed Sampling, is an emerging technique, which has shown that sparse or compressible signal can be reconstructed from far fewer measurements than what is considered by Nyquist theorem \cite{rj9}. Compressed sensing theory usually consists of sparse representation, measurement matrix, and sparse coefficient reconstruction.

\subsubsection{Sparse representation}
 Consider a $N \times 1$ signal ${\bm{f}} = {\left[ {f_1,f_2, \cdots ,f_N} \right]^{\rm T}}$, which can be expressed as a linear combination of a complete basis ${\bm{\Psi }}$:
\begin{equation}
{\bm{f}} = \sum\limits_{i = 0}^N {{c_i}} {\psi _i} = {\bm{\Psi c}}
\label{eq.5}
\end{equation}
where ${\bm{c}} = {\left[ {{c_1},{c_2}, \cdot  \cdot  \cdot ,{c_N}} \right]^{\rm T}}$, ${{\bm{\Psi }}} = \left[ {{\bm{\psi} _1},{\bm{\psi} _2}, \cdot  \cdot  \cdot ,{\bm{\psi} _N}} \right]$. If there are $K$ non-zero elements in $\bm{c}$, then $\bm{c}$ is called $K$-sparse and we say $\bm{f}$ is sparse in the ${\bm{\Psi }}$ domain.

\subsubsection{Measurement matrix}
$M$ projections of the signal $\bm{f}$ constitute a $M \times N(M \ll N)$ matrix ${\bm{\Phi }}$ called measurement matrix, then the sampling process can be described as:

\begin{equation}
{\bm{y}} = {\bm{\Phi f}} = {\bm{\Phi }}{{\bm{\Psi }}}{\bm{c}} = {\bm{Ac}}
\label{eq.6}
\end{equation}
where ${\bm{A}} = {\bm{\Phi }}{{\bm{\Psi }}}$ is known as measurement matrix. It is clear that the process of recovering $\bm{c}$ from $\bm{y}$ is an ill-posed problem because $M \ll N$ and seems impossible to solve such underdetermined linear equation. However, if the sparse coefficients $\bm{c}$ is $K$-sparse, and $\bm{\Phi }$ satisfy the Restricted Isometry Property (RIP) criterion \cite{rj21}, then $\bm{c}$ can be accurately recovering with a high probability from $M$ measurements. The RIP criterion can be expressed as

\begin{equation}
(1 - {\delta _K})||{\bm{c}}||_2^2 \le ||{{\bm{\Psi }}}{\bm{c}}||_2^2 \le (1 + {\delta _K})||{\bm{c}}||_2^2.
\label{eq.7}
\end{equation}
The restricted isometry constant ${\delta _K}(0 < {\delta _K} < 1)$ is defined as the smallest constant for which this property holds for all $K$-sparse vectors $\bm{c}$.

\subsubsection{Sparse coefficients reconstruction algorithm}
As can be seen from above, the sparse coefficient vector $\bm{c}$ can be recovered with high probability from $\bm{y}$ by solving the following  ${\ell _0}$-norm optimization:

\begin{equation}
\arg {\min _{\bm{c}}}{\left\| {{\bm{y}} - {\bm{A c}}} \right\|_2}{\rm{  }}\quad s.t.\quad{\rm{  }}{\left\| {\bm{c}} \right\|_0}{\rm{ =  }}K{\rm{ }}
\label{eq.8}
\end{equation}
where ${\ell _0}$-norm, $||{\bf{x}}|{|_0}$, denote the number of non-zero entries in the argument and $||{\bf{x}}|{|_2}$ indicates its Euclidean norm. Some iterative greedy algorithms have been developed to solve such combinatorial optimization problem, such as Iterative hard thresholding (IHT) \cite{rj17}, Subspace Pursuit (SP) \cite{rj20}, and a series of algorithms derived from them.  

\section{Sparse representation of the radiation flux}

\begin{figure}
\centering
{\includegraphics[width=15 cm]{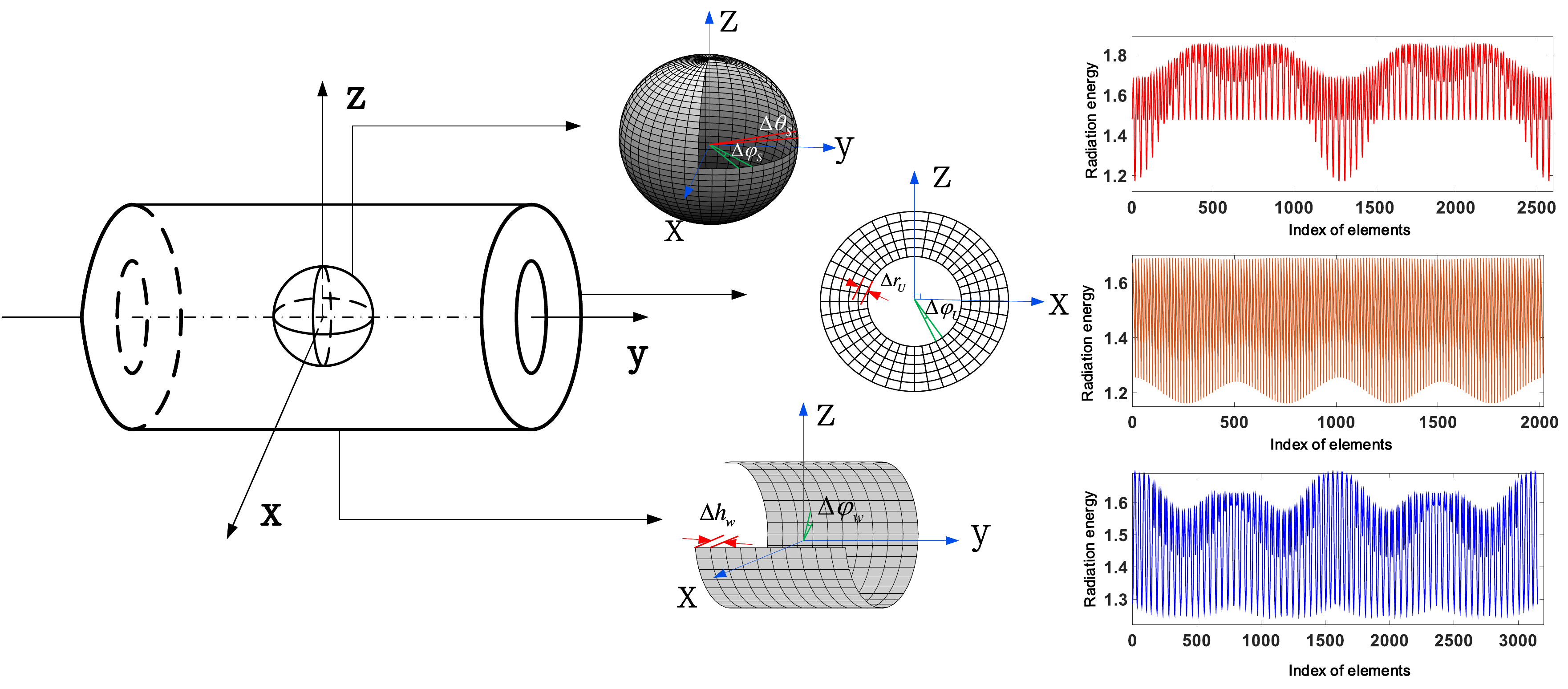}}
\caption{Division of radiation surfaces in cylinder model in cartesian coordinate}\label{fig2}
\end{figure}

In the cylindrical cavity model of ICF, as shown in Figure \ref{fig2}, the radiation region located in the cavity consists of three parts, i.e., spherical surface of the capsule, the bottom and the top surface and the side surface of the cylindrical cavity. Here, $B_s$, $B_u$ and $B_w$ are used to represent the radiation flux of theses three parts respectively and the total radiation flux can be expressed as $B=B_{S}+B_{u}+B_{w}$. Since the X-ray converted from incident laser are uniformly and continuously distributed on the radiation regions, the radiation flux could be represented by linear combinations of some complete orthogonal polynomials. In this section, three orthogonal bases which are constructed by three sets of orthogonal polynomials are used to sparsely represent the radiation flux.

\subsection{Sparse representation of the radiation flux on capsule}

\begin{figure}  
\centering  
\includegraphics[width=0.55\textwidth]{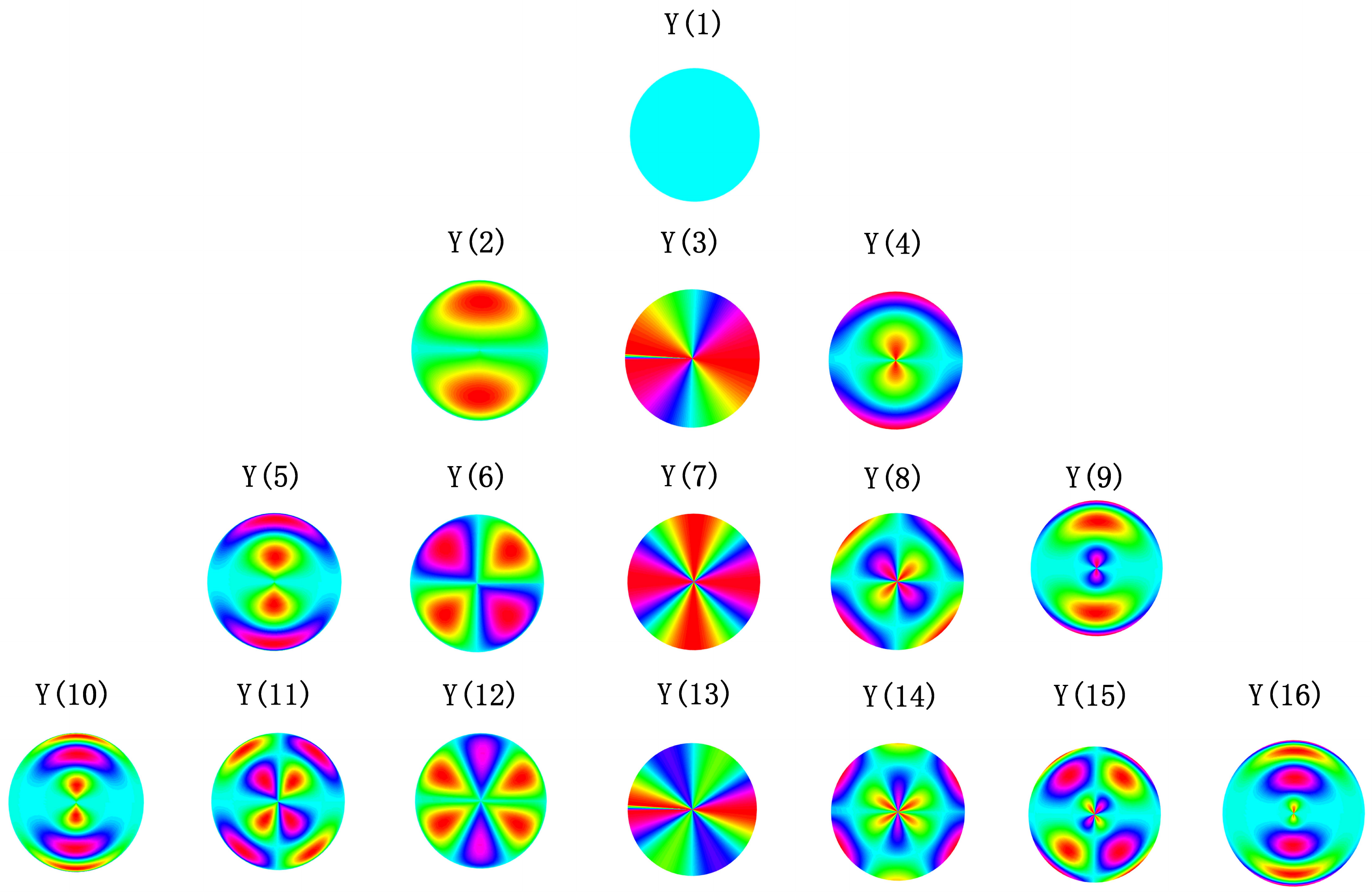}    
\caption{3D rendering of the first 16 terms of spherical harmonic polynomials} 
\label{fig3}
\end{figure}

Let $\left(\theta_{s}^{(i)}, \varphi_{s}^{(i)}\right)$ be a spherical coordinate of the $i$-th element, then the radiation $B_{s}\left(\theta_{s}^{(i)}, \varphi_{s}^{(i)}\right)$ can be represented by a linear combinations of the spherical harmonics as follows \cite{rj15}:

\begin{equation} 
B_{s}\left(\theta_{s}^{(i)}, \varphi_{s}^{(i)}\right)=\sum_{n_{s}=0}^{+\infty} c_{n_{s}} Y_{\left(n_{s}\right)}\left(\theta_{s}^{(i)}, \varphi_{s}^{(i)}\right),
\label{eq.9}
\end{equation}
where the subscript $n_{s}$ is a polynomial ordering index starting with  ${n_s} = 0$, ${c_{{n_s}}}$ is its coefficient and $Y({\theta}_s,{\varphi}_s )$ is the spherical harmonics which can be represented as


\begin{equation} 
Y_{{m_S}}^{{k_S}}({\theta _s},{\varphi _s}) = N_{{m_S}}^{{k_S}}P_{{m_S}}^{{k_S}}(\cos {\theta _s})\left\{ {\begin{array}{*{20}{c}}
   {\cos {k_s}{\varphi _s}{\rm{  }} \quad if \quad {\rm{ }}{k_S} \ge 0}  \\
   {\sin {k_s}{\varphi _s}{\rm{   }} \quad if \quad {\rm{ }}{k_S} < 0}  \\
\end{array}}\right.{\rm{,  }}\quad( - {m_S} \le {k_S} \le {m_S})
\label{eq.10}
\end{equation}

where $N_{m_{S}}^{k_{S}}=\sqrt{\frac{\left(\left(2 m_{s}+1\right)\left(m_{s}-k_{s}\right) !\right)}{\left(4 \pi\left(m_{s}+k_{S}\right) !\right)}}$ is the normalization constant and $P_{m_{s}}^{k_{s}}\left(\cos \theta_{s}\right)$ is the Associated Legendre polynomial.
As shown in Figure \ref{fig2}, let $\Delta {\theta}_s $ and $\Delta {\varphi}_s $ be the minimal angles of subdividing the capsule over the domain $\Omega ({\theta}_s ,{\varphi}_s ) = [0,\pi ] \times [0,2\pi )$. Equation \eqref{eq.10} can be transformed into an approximate form as 

\begin{equation} 
B_{S}\left(\theta_{s}, \varphi_{s}\right)=\sum_{j=1}^{N_{s}} \sum_{n_{s}=0}^{+\infty} c_{n_{s}} Y_{n_{s}}\left(\theta_{s}^{(j)}, \varphi_{s}^{(j)}\right) \approx \sum_{j=1}^{N_{s}} \sum_{n_{s}=0}^{L_{s}} c_{n_{s}} Y_{n_{s}}\left(\theta_{s}^{(j)}, \varphi_{s}^{(j)}\right)=\bm{Y} \bm{c}^{s},
\label{eq.11}
\end{equation}

where ${L_S}$ is the number of spherical harmonics expansion terms, $\bm{c}^{s}=\left[c_{1}, c_{2}, \cdots, c_{L_{S}}\right]^{\mathrm{T}}$ is coefficient vector and $\bm{Y}$ is spherical harmonics basis which can be expressed as

\begin{equation}
{{\bm{Y}}_{{N_S} \times {L_S}}} = \left[ {\begin{array}{*{20}{c}}
   {{Y_0}(\theta _s^{(1)},\varphi _s^{(1)})} & {{Y_1}(\theta _s^{(1)},\varphi _s^{(1)})} &  \cdots  & {{Y_{{L_S}}}(\theta _s^{(1)},\varphi _s^{(1)})}  \\
   {{Y_0}(\theta _s^{(1)},\varphi _s^{(2)})} & {{Y_1}(\theta _s^{(1)},\varphi _s^{(2)})} &  \cdots  & {{Y_{{L_S}}}(\theta _s^{(1)},\varphi _s^{(2)})}  \\
    \vdots  &  \vdots  &  \ddots  &  \vdots   \\
   {{Y_0}(\theta _s^{({N_\theta })},\varphi _s^{({N_{{\varphi _s}}})})} & {{Y_1}(\theta _s^{({N_\theta })},\varphi _s^{({N_{{\varphi _s}}})})} &  \cdots  & {{Y_{{L_S}}}(\theta _s^{({N_\theta })},\varphi _s^{({N_{{\varphi _s}}})})}  \\
\end{array}} \right],
\label{eq.12}
\end{equation}
where $N_{{\theta}_s}=\frac{\pi}{\Delta \theta_{s}}$ and $N_{\varphi_{s}}=\frac{2 \pi}{\Delta \varphi_{s}}$ denote the number of elements divided along the longitude and latitude respectively, and ${N_S}$ is the total number of discrete elements on the capsule which can be calculated by $N_{S}=N_{\theta} \times N_{\varphi}$. The 3D pictures of the first 16 terms of the spherical harmonics are shown in Figure \ref{fig3}.

\subsection{Annular Zernike Polynomials}

The bottom and the top surface of the cylindrical cavity is annular because it has two laser entrance holes. As described in \cite{rj22}, Zernike annular polynomials, defined on an annular region, is obtained from the Zernike circle polynomials \cite{rj22} by using the Gram-Schmidt orthogonalization process. The Zernike annular polynomials are widely used to mathematically describe wavefront aberrations of optical systems due to it is orthogonal over an annular domain, and it can uniquely describe any continuous function with the same definition domain \cite{rj21}. 
Let $\left(r_{u}^{(i)}, \varphi_{u}^{(i)}\right)$ be a polar coordinate of the $i$-th element of the bottom surface, the annular Zernike polynomials have the following forms in a polar coordinate:

\begin{equation} 
U_{l_{u}}^{k_{u}}\left(r, \tilde r ; \varphi_{u}\right)=\left\{\begin{array}{ll}{R_{l u}^{k_{u}}(r, \tilde r) \cos k_{u} \varphi_{u}} & {\text { for } \quad k_{u} \geq 0} \\ {R_{l_{u}}^{k_{u}}(r, \tilde r) \sin k_{u} \varphi_{u}} & {\text { for } \quad k_{u}<0}\end{array}\right.
\label{eq.13}
\end{equation}
where $r(0 \leq r \leq 1)$ and $\tilde r(\tilde r \leq r \leq 1)$ are the outer and inner semi-diameter of the annular region respectively, $\varphi_{u}\left(0 \leq \varphi_{u} \leq 2 \pi\right)$ is the circumferential angle, $l_{u} \geq 0$, $k_{u} \geq 0$ are integers, $l_{u}-\left|k_{u}\right| \geq 0$ and even, $R_{l_{u}}^{k_{u}}(r, \tilde r)$ is the radial components of the polynomials. The orthonormality of the annular Zernike polynomials can be expressed as

\begin{equation}
{{\int\limits_{\tilde r} ^1 {\int\limits_0^{2\pi } {U_{{l_u}}^{{k_u}}(r,\tilde r ;{\varphi _u})U_{l{'_u}}^{k{'_u}}(r,\tilde r ;{\varphi _u})} } rdrd{\varphi _u}} \mathord{\left/
 {\vphantom {{\int\limits_{\tilde r} ^1 {\int\limits_0^{2\pi } {U_{{l_u}}^{{k_u}}(r,\tilde r ;{\varphi _u})U_{l{'_u}}^{k{'_u}}(r,{\tilde r} ;{\varphi _u})} } rdrd{\varphi _u}} {\int\limits_\varepsilon ^1 {\int\limits_0^{2\pi } {rdrd{\varphi _u} = {\delta _{{l_u}l{'_u}}}} } }}} \right.
 \kern-\nulldelimiterspace} {\int\limits_\varepsilon ^1 {\int\limits_0^{2\pi } {rdrd{\varphi _u} = {\delta _{{l_u}l{'_u}}}} } }}{\delta _{{k_u}k{'_u}}},
\label{eq.14}
\end{equation}
where $\delta_{l_{u} l^{\prime}_{u}}\left(\delta_{k_{u} k_{u}^{\prime}}\right)$ stands for the Kronecker delta, equal to one if $l_{Z}=l_{Z}^{\prime}\left(k_{Z}=k_{Z}^{\prime}\right)$ and to zero otherwise.
when $\tilde r = 0$, the annular polynomials reduce to the Zernike circle polynomials. Let $l_{u}=2 j+k_{u}$, the radial polynomials expression of the annular Zernike polynomials can be obtained by the recurrence formula: \cite{rj22}
when $k_{u}=0$

\begin{equation} 
R_{2 j}^{0}(r, \tilde r)=R_{2 j}^{0}\left(\frac{r^{2}-{\tilde r}^{2}}{1-{\tilde r}^{2}}\right)^{1 / 2}=P_{j}\left[\frac{2\left(r^{2}-{\tilde r}^{2}\right)}{1-{\tilde r}^{2}}-1\right],
\label{eq.15}
\end{equation}
where $P_{j}(\cdot)$ is the Legendre polynomials.
When $k_{u}>0$, the radial polynomials are given by the recurrence relationship

\begin{equation} 
R_{2 j+k_{n}}^{k_{u}}(r, \tilde r)=\left[\frac{1-{\tilde r}^{2}}{2\left(2 j+k_{u}+1\right) H_{j}^{k_{u}}}\right]^{1 / 2} r^{k_{i}} Q_{j}^{k_{u}}\left(r^{2}\right),
\label{eq.16}
\end{equation}
where $ Q_{j}^{k_{u}}(\tau)$ is a set of orthogonal polynomials obtained by orthogonalizing the sequence $1, \tau, \cdots, \tau^{j}$ over the interval $\left({\tilde r}^{2}, 1\right)$ with a weight function $\tau^{k_{u}}$, $ H_{j}^{k_{u}}$ are the normalization constants. the recurrence formula of the $Q_{j}^{k_{u}}(\tau)$ and $H_{j}^{k_{u}}$ as follows:

\begin{equation} 
Q_{j}^{k_{u}}\left(r^{2}\right)=\frac{2\left(2 j+2 k_{u}-1\right)}{\left(j+k_{u}\right)\left(1-{\tilde r}^{2}\right)} \frac{H_{j}^{k_{u}-1}}{Q_{j}^{{k_u}-1}(0)} \sum_{i=0}^{j} \frac{Q_{i}^{{k_u}-1}(0) Q_{i}^{{k_u}-1}\left(r^{2}\right)}{H_{i}^{{k_u}-1}}
\label{eq.17}
\end{equation}

\begin{equation}
H_j^{{k_u}} =  - \frac{{2(2j + 2{k_u} - 1)}}{{(j + {k_u})(1 - {{\tilde r} ^2})}}\frac{{Q_{j + 1}^{{k_u} - 1}(0)}}{{Q_j^{{k_u} - 1}(0)}}H_j^{{k_u} - 1}
\label{eq.18}
\end{equation}
especially, when $k_{u}=0$,

\begin{equation}
Q_j^0({r^2}) = R_{2j}^0(r,\tilde r );{\rm{   }}H_j^0 = \frac{{1 - {{\tilde r} ^2}}}{{2(2j + 1)}}{\rm{ }}.
\label{eq.19}
\end{equation}
When $k_{u}<0$, the radial polynomials can be obtained by the relationship:

\begin{equation} 
R_{2 j+k_{u}}^{k_{u}}(r, \tilde r)=R_{2 j+k_{u}}^{\left|k_{u}\right|}(r, \tilde r).
\label{eq.20}
\end{equation}
When $ k_{u}=l_{u}$

\begin{equation}
R_{{k_u}}^{{k_u}}(r,\tilde r ) = \frac{{{r^{{k_u}}}}}{{{{(\sum\limits_{i = 0}^{{k_u}} {{r^{2i}}} )}^{1/2}}}}.
\label{eq.21}
\end{equation}

\begin{figure}  
\centering  
\includegraphics[width=0.55\textwidth]{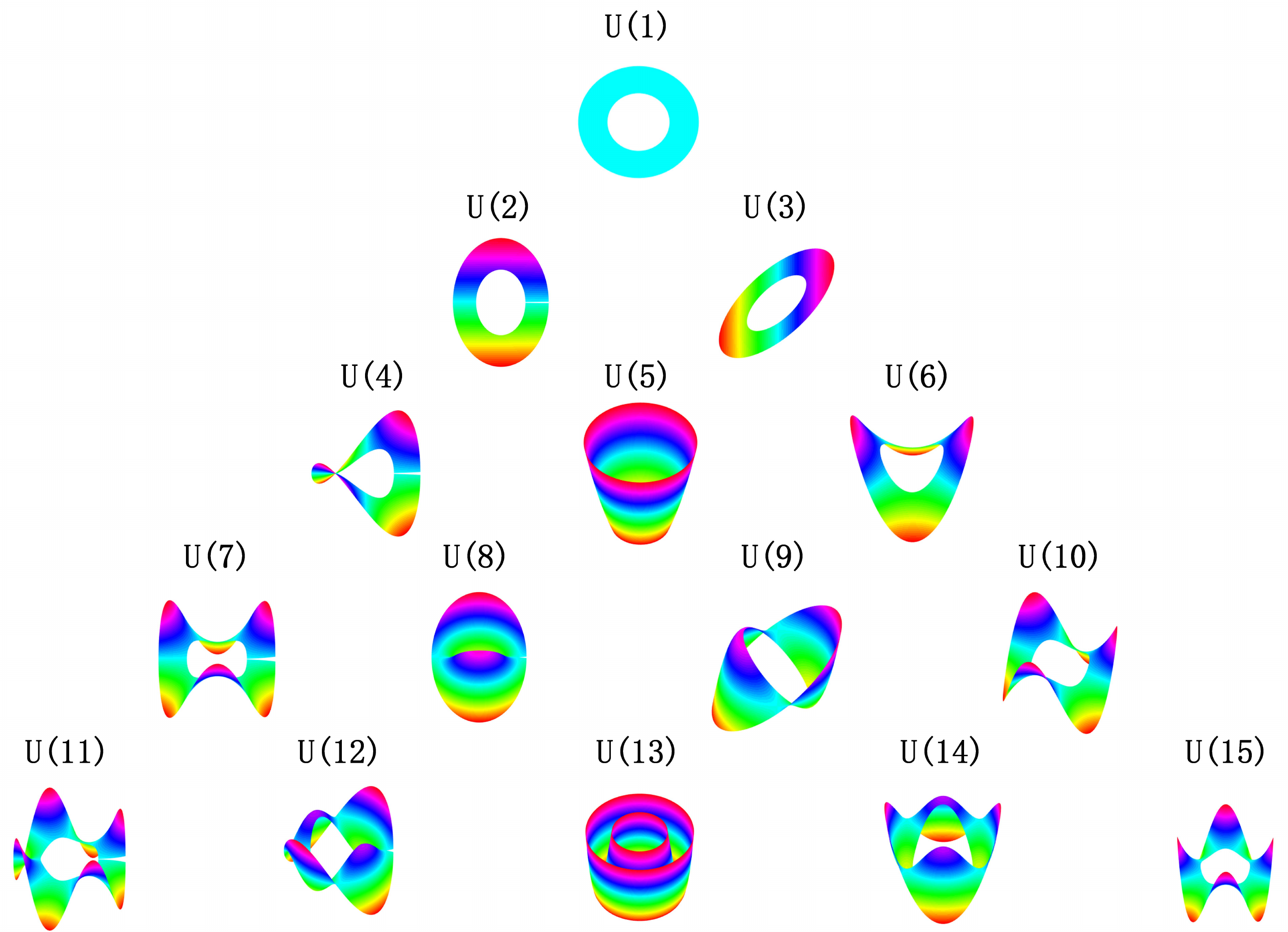}    
\caption{3D rendering of the first 15 terms of Annular Zernike polynomials} 
\label{fig4}
\end{figure}

In practice, the outer diameter of the bottom surface is not equal to 1, so we require to map the reality interval $\left[r_{\text {outer}}, r_{\text {inner}}\right]$ to the interval $[\tilde r, 1]$. Similar to the radiation flux representation on the capsule, the radiation flux $B_{u}\left(r_{u}^{(i)}, \varphi_{u}^{(i)}\right)$ can be expressed by a linear combinations of the annular Zernike polynomials as

\begin{equation} 
B_{u}\left(r, \varphi_{u}\right)=\sum_{i=1}^{N_{u}} \sum_{n_{u}=0}^{+\infty} c_{n_{u}} U_{n_{u}}\left(r^{(i)}, \tilde r ; \varphi_{u}^{(i)}\right) \approx \sum_{j=1}^{N_{u}}
 \sum_{n_{u}=0}^{L_{u}} c_{n_{u}} U_{n_{u}}\left(r^{(i)}, \tilde r ; \varphi_{u}^{(i)}\right)=\bm{U} \bm{c}^{u},
\label{eq.22}
\end{equation}
where $n_{u}$ is a polynomial ordering index starting with $n_{u}=0$,  $\boldsymbol{U}$ is the annular Zernike polynomials with matrix form which presented as

\begin{equation}
{{\bm{U}}_{{N_u} \times {L_u}}} = \left[ {\begin{array}{*{20}{c}}
   {{U_0}({r^{(1)}},\tilde r ;\varphi _u^{(1)})} & {{U_1}({r^{(1)}},\tilde r ;\varphi _u^{(1)})} &  \cdots  & {{U_{{L_u}}}({r^{(1)}},\tilde r ;\varphi _u^{(1)})}  \\
   {{U_0}({r^{(1)}},\tilde r ;\varphi _u^{(2)})} & {{U_1}({r^{(1)}},\tilde r ;\varphi _u^{(2)})} &  \cdots  & {{U_{{L_u}}}({r^{(1)}},\tilde r ;\varphi _u^{(2)})}  \\
    \vdots  &  \vdots  &  \ddots  &  \vdots   \\
   {{U_0}({r^{({N_r})}},\tilde r ;\varphi _u^{({N_{{\varphi _u}}})})} & {{U_1}({r^{({N_r})}},\tilde r ;\varphi _u^{({N_{{\varphi _u}}})})} &  \cdots  & {{U_{{L_u}}}({r^{({N_r})}},\tilde r ;\varphi _u^{({N_{{\varphi _u}}})})}  \\
\end{array}} \right]
\label{eq.23}
\end{equation}

where the subscript $N_{r}$ and $N_{\varphi_{u}}$ are the number of elements divided along the radial direction and rotation direction respectively. $N_{u}$ is the total number of elements of one of the end face of the cylinder cavity which can be calculated by $N_{u}=N_{r} \times N_{\varphi_{u}}$. $L_{u}$ is the number of the Zernike annular polynomials expansion terms which can be obtained by the formula $L_{u}=\left(n_{u}+2\right)\left(n_{u}+1\right) / 2$. The 3D pictures of the first 15 terms of the annular Zernike polynomials are shown in Figure \ref{fig4}.

\subsection{Legendre Fourier polynomials}

Since the side of the cylindrical cavity is a cylindrical surface, it is no doubt that cylindrical coordinates $\left(\rho, \varphi_{w}, z\right)$ seem to be the optimal choice for describing $B_{w}$, where $\rho$, $\varphi_{w}$ and $Z$ denote radius, azimuth angle and height of the cylindrical coordinate. Obviously, $B_{w}$ is only related to azimuth angle and height and not to radius.

According to the construction principle of multivariable orthogonal polynomials, two sets of one-dimensional polynomials which are defined in the azimuth direction and height direction respectively, can be employed to construct a two-dimensional orthogonal polynomials for describing $B_{w}$. Since the cylinder surface profile along the azimuthal direction is closed and periodic, a Fourier series is suitable for characterizing the azimuthal coordinate, which can be expressed as

\begin{equation} 
F_{k_{w}}\left(\varphi_{w}\right)=\left\{\begin{array}{ll}{\cos k_{w} \varphi_{w}} & {k_{w} \geq 0} \\ {\sin k_{w} \varphi_{w}} & {k_{w}<0}\end{array}\right.
\label{eq.24}
\end{equation}
where $k_{w}=0, \pm 1, \pm 2, \cdots$. 

In height direction, coordinate $z$ is restricted in the interval $[-h, h]$, where $h$ is half the height of the cylindrical cavity. A nature choice for $z$ coordinate is a Fourier series \cite{rj24}. Similar to the Associated Legendre polynomials used for spherical harmonics, some one-dimensional-polynomials such as Legendre polynomials and Chebyshev polynomials are suitable for the $z$ coordinate. Since Legendre polynomials are orthogonal across the interval $[-1,1]$, which are conveniently mapped to the interval $[-h, h]$, here we consider use Legendre polynomials for characterizing the $z$ coordinate. Then a two-dimensional polynomials can be defined as the tensor products of Fourier series and Legendre polynomials in the variables $z$ and $\varphi_{w}$:

\begin {equation} 
W_{n_{w}}\left(z, \varphi_{w}\right)=P_{l_{w}}(z) F_{k_{w}}\left(\varphi_{w}\right),
\label{eq.25}
\end{equation}
where $P_{l_{w}}(z)$ is Legendre polynomials with $l_{w}$ degree,  $W_{n_{w}}\left(z, \varphi_{w}\right)$ termed Legendre Fourier (LF) polynomials defined in the domain $\Omega=\left\{\left(z, \varphi_{w}\right) \in[0,2 \pi] \times[-1,1]\right\}$, $n_{w}$ denotes the ordering index of the LF polynomials starting with $n_{w}=0$. An integer $n_{w}$ represents the degree of the LF polynomials which can be calculated as $n_{w}=l_{w}+\left|k_{w}\right|$, where the operator $|\cdot|$ means taking the absolute value. It is clearly that the LF polynomials are separable in the cylindrical coordinates  $z$ and $\varphi_{w}$.

\begin{figure}  
\centering  
\includegraphics[width=0.55\textwidth]{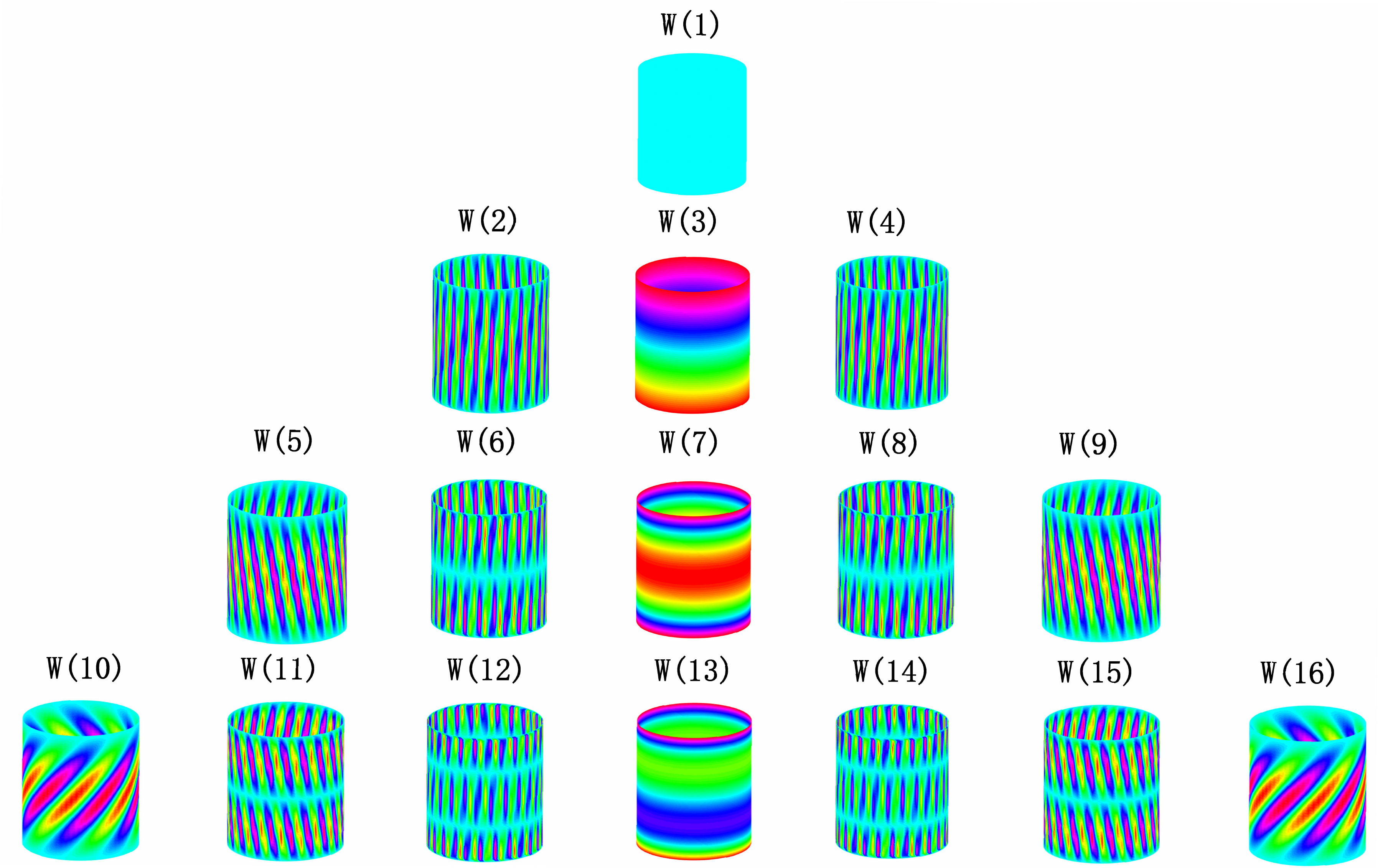}    
\caption{3D rendering of the first 16 terms of Legendre-Fourier polynomials} 
\label{fig5}
\end{figure}

$B_{w}\left(z, \varphi_{w}\right)$ can be approximately presented by the LF polynomials as

\begin {equation}
{B_w}(z,{\varphi _w}) = \sum\limits_{i = 1}^{{N_w}} {\sum\limits_{{n_w} = 0}^{ + \infty } {{c_{{n_w}}}{W_{{n_w}}}({z^{(i)}},\varphi _w^{(i)})} }  \approx \sum\limits_{j = 1}^{{N_w}} {\sum\limits_{{n_w} = 0}^{{L_w}} {{c_{{n_w}}}{W_{{n_w}}}({z^{(i)}},\varphi _w^{(i)})} }  = {\bm{W}}{{\bm{c}}^w},
\label{eq.26}
\end{equation}
where $c_{n_{w}}$ is the expansion coefficient of the LF polynomials.$\boldsymbol{W}$ denotes the LF orthogonal basis which can be written as

\begin{equation}
{{\bm{W}}_{{N_w} \times {L_w}}} = \left[ {\begin{array}{*{20}{c}}
   {{W_0}({z^{(1)}},\varphi _w^{(1)})} & {{W_1}({z^{(1)}},\varphi _w^{(1)})} &  \cdots  & {{W_{{L_w}}}({z^{(1)}},\varphi _w^{(1)})}  \\
   {{W_0}({z^{(1)}},\varphi _w^{(2)})} & {{W_1}({z^{(1)}},\varphi _w^{(2)})} &  \cdots  & {{W_{{L_w}}}({z^{(1)}},\varphi _w^{(2)})}  \\
    \vdots  &  \vdots  &  \ddots  &  \vdots   \\
   {{W_0}({z^{({N_z})}},\varphi _w^{({N_{{\varphi _w}}})})} & {{W_1}({z^{({N_z})}},\varphi _w^{({N_{{\varphi _w}}})})} &  \cdots  & {{W_{{L_w}}}({z^{({N_z})}},\varphi _w^{({N_{{\varphi _w}}})})}  \\
\end{array}} \right],
\label{eq.27}
\end{equation}

where the subscript $N_{z}$ and $N_{\varphi_{w}}$ are the number of elements divided along the radial direction and rotation direction respectively. $N_{w}$ is the total number of elements of the side surface of the cylindrical cavity which can be calculated by $N_{w}=N_{z} \times N_{\varphi_{w}}$. $L_{w}$ is the number of the LT polynomials expansion terms. The top 9 expansion terms of the LT polynomials are given in Table \ref{table1}. The 3D pictures of the first 16 terms of the LF polynomials are shown in Figure \ref{fig5}.

\begin{table}
\caption{The first 9 expansion terms of the LF polynomials sequence}
\label{table1}
\begin{center}
\begin{tabular}{cccc}
\toprule
 Terms & Order & Polynomials & Expression \\
\midrule
 ${W_0}$ & 0 & ${L_0}{{F}_0}$ & 1  \\
 ${W_1}$ & 1 & ${L_0}{{F}_{-1}}$ & $\sin {k_{w}}\varphi_{w}$  \\
 ${W_2}$ & 1 & ${L_0}{{F}_1}$ & $\cos {k_{w}}\varphi_{w}$ \\
 ${W_3}$ & 1 & ${L_1}{{F}_0}$ & $z$  \\
 ${W_4}$ & 2 & ${L_0}{{F}_{-2}}$ & $\sin {2{k_{w}}\varphi_{w}}$  \\
 ${W_5}$ & 2 & ${L_0}{{F}_2}$ & $\cos {2{k_{w}}\varphi_{w}}$  \\
 ${W_6}$ & 2 & ${L_1}{{F}_{-1}}$ & $z\sin {k_{w}}\varphi_{w}$  \\
 ${W_7}$ & 2 & ${L_1}{{F}_1}$ & $z\cos {k_{w}}\varphi_{w}$  \\
 ${W_8}$ & 2 & ${L_2}{{F}_0}$ & ${{(3{z^2} - 1)} \mathord{\left/
 {\vphantom {{(3{z^2} - 1)} 2}} \right.
 \kern-\nulldelimiterspace} 2}$ \\
\bottomrule
\end{tabular}
\end{center}
\end{table}

\subsection{Sparse Representation Model of Radiation Flux}
From Subsection 3.1-3.3, the total radiation flux in the cavity can be expressed in the form of sum:

\begin{equation}
B = {B_s} + {B_u} + {B_w} = {{\bm{Y}}_{{N_s} \times {L_s}}}{{\bm{c}}^s} + {{\bm{U}}_{{N_u} \times {L_u}}}{{\bm{c}}^u} + {{\bm{W}}_{{N_w} \times {L_w}}}{{\bm{c}}^w},
\label{eq.28}
\end{equation}
or matrix form:

\begin{equation}
{\bm{B}} = \left[ {\begin{array}{*{20}{c}}
   {{{\bm{Y}}_{{N_s} \times {L_s}}}} & {} & {}  \\
   {} & {{{\bm{U}}_{{N_u} \times {L_u}}}} & {}  \\
   {} & {} & {{{\bm{W}}_{{N_w} \times {L_w}}}}  \\
\end{array}} \right] \times \left[ {\begin{array}{*{20}{c}}
   {{{\bm{c}}^s}}  \\
   {{{\bm{c}}^u}}  \\
   {{{\bm{c}}^w}}  \\
\end{array}} \right] = {\bm{\Psi c}}.
\label{eq.29}
\end{equation}
Equation \eqref{eq.29} can be written as a functional form

\begin{equation} 
f({\bm{c}}) = ({\bm{I}} - {\bm{V}})({\bm{\Psi c}}) + C({\bm{\Psi c}}){\circ ^{{1 \mathord{\left/
 {\vphantom {1 \beta }} \right.
 \kern-\nulldelimiterspace} \beta }}} - {\bm{E}}
\label{eq.30}
\end{equation}
which has a unique unknown variable, i.e., sparse coefficient $\bm{c}$. 

In this section, three sets of orthogonal polynomials are used to represent the radiation flux in a cylindrical cavity, and the radiation energy balance equation is transformed into an equation about the sparse coefficients.

\section{Radiation flux reconstruction under Compressive Sensing framework}

\subsection{Compressed Observation Model of Radiation Flux}
Compressed Sensing is usually works on linear system. First-order Taylor expansion around a point approximation is usually applied to linearization in nonlinear CS that is similar to the case in Equation \eqref{eq.30}. Equation \eqref{eq.30} can be expanded at point $c^{*}$ as

\begin{equation} 
f({\bm{c}}) \approx f({{\bm{c}}^ \star }) + {f_{{{\bf{c}}^ \star }}}({\bm{c}} - {{\bm{c}}^ \star }),
\label{eq.31}
\end{equation}
where $f_{c^{\star}}$ is a linear operator such as Jacobian matrix which can be expressed as

\begin{equation} 
{f_{{{\bm{c}}^ \star }}} = ({\bm{I}} - {\bm{V}}){\bm{\Psi }} + {\mathop{\rm diag}\nolimits} \left[ {\frac{1}{\beta }C({\bm{\psi }}{{\bm{c}}^ \star }){\circ ^{{{(1} \mathord{\left/
 {\vphantom {{(1} {\beta  - 1}}} \right.
 \kern-\nulldelimiterspace} {\beta  - 1}})}}} \right],
\label{eq.32}
\end{equation}
where $\operatorname{diag}(\bullet)$ denotes a diagonal matrix whose elements on the diagonal are composed of $\bullet$. Then Equation \eqref{eq.32} can be rewritten as:

\begin{equation} 
{f_{{{\bf{c}}^ \star }}}{\bf{c}} = {f_{{{\bf{c}}^ \star }}}{{\bf{c}}^ \star } - f({{\bf{c}}^ \star }).{\rm{ }}
\label{eq.33}
\end{equation}

As the experimental results shown in Table 2, in the case of smaller grid, the computation is very large for solving view-factor matrix via Newton-Raphson method. In accordance to the CS theory, a measurement matrix of $M \times N(M \ll N)$ can be used for obtaining linear and non-adaptive measurements from Equation \eqref{eq.33}, which can be expressed as

\begin{equation}
{\bm{\Phi }}{f_{{{\bm{c}}^ \star }}}{\bm{c}} = {\bm{\Phi }}({f_{{{\bm{c}}^ \star }}}{{\bm{c}}^ \star } - f({{\bm{c}}^ \star })).{\rm{ }}
\label{eq.34}
\end{equation}

The commonly used measurement matrices have stochastic Gaussian matrices, random Bernoulli matrices, partial orthogonal matrices, structured random matrices and cyclic matrices and sparse random matrices \cite{rj10}. Since the basis $\boldsymbol{\Psi}$ is orthogonal, from which the randomly sampled elements must meet the orthogonal, and also to meet the Restricted Isometry Property (RIP), the sampling points above the elements can be directly selected. This means only needs to calculation the sampling elements composed of measurement matrix. In practical, the measurement matrix is a similar diagonal matrix consisting of the sampling of row index of $f_{c^{\star}}$.

Let $A=\boldsymbol{\Phi} f_{c^{\star}}$, $\boldsymbol{y}=\boldsymbol{\Phi}\left(f_{c^{\star}} c^{\star}-f\left(\boldsymbol{c}^{\star}\right)\right)$, Equation \eqref{eq.34} can be written as

\begin{equation}
{\bm{Ac}} = {\bm{y}}.
\label{eq.35}
\end{equation}

Since $M<N$, in general, recovery of $\bm{c}$ is a combination problem which is known to be NP-hard. Fortunately, under stricter conditions on $\bm{A}$ (i.e., Restricted Isometry Property), a range of different algorithms can be used to recover $\bm{c}$ from roughly $O(K \log (N))$ measurements \cite{rj9}. Based on the results of Section 3, we can roughly estimate the number of samples for capsule, the bottom and the top surface of the cavity and the side of the cavity, respectively. For example, for a model with a total elements number of 9776 (where the the number of elements for capsule, the bottom and top surface of the cavity, and the side of the cavity are 2592, 2016, 2016 and 3152, respectively), the number of samples for capsule is $30 \times \log (2592)=102.4$, the bottom and top surface have the same samples of $35 \times \log (2016)=115.6$, and the number of samples on the side of cavity is $100 \times \log (3152)=349.9$, and the total number of samples can be estimated as $150+150 \times 2+400=850$. The number of samples in other models is shown in Table \ref{tab3}.

\subsection{Algorithms for solving model}

Sparse coefficient $\bm{c}$ can be recovered by optimizing the following sparse constraint problems

\begin{equation}
{\bm{\hat c}} = \arg \min {\left\| {{\bm{y}} - {\bm{Ac}}} \right\|_2}{\rm{    }}\quad s.t.\quad {\rm{    }}{\left\| {\bm{c}} \right\|_0} \le K
\label{eq.36}
\end{equation}
some greedy iterative algorithms such as hard thresholding algorithms and orthogonal matching algorithms can be used for solving Equation \eqref{eq.36}. In this section, we proposed a new iterative greedy algorithm for solving the radiation flux compressed observation model based on two typical CS reconstruction algorithms.

\subsubsection{Conjugate Gradient Iterative Hard Thresholding Algorithm}
Iterative hard thresholding (IHT) algorithm is usually used for CS reconstruction due to its simple operation and low computational complexity, and it's iterative formula as 

\begin{equation}
{{\bm{c}}_{n + 1}} = {H_K}[{{\bm{c}}_n} + \mu {{\bm{A}}^{\rm T}}({\bm{y}} - {\bm{A}}{{\bm{c}}_n})],
\label{eq.37}
\end{equation}
where $H_{K}(\bullet)$ is a nonlinear operator that sets all but the absolute value of largest $K$ elements of $\bm{c}$ to zero, $\mu $ is the step size and usually taken as a constant in IHT algorithm. The convergence of this algorithm was proven in [dd] under the condition that  $\|\bm{A}\|_{2}<1$, in which case, the IHT algorithm converges to a local minimum. Since step size and the scale of sensing matrix have a great influence on the convergence of IHT \cite{rj17}, Normalized Iterative Hard Thresholding algorithm (NIHT) was proposed in to improve IHT algorithm \cite{rj18}. The restriction of the snesing matrix $\bm{A}$ in NIHT is replaced by a descending iteration factor, which greatly improves the convergence and stability. However, the search direction in NIHT algorithm is gradient direction, it may result slow convergence near the minimum. Conjugate Gradient Iterative Hard Thresholding (CGIHT) algorithm was proposed in \cite{rj19}, in which, conjugate gradient direction was adopted as the search direction to speed up the convergence. The CGIHT algorithm can be stated as Algorithm \ref{alg1}, in which, $\emptyset $ denotes an empty set and $\bm{d}$ denotes search direction, $\gamma$ is an orthogonalization weight, the support of $\bm{c}$ denoted by $T_{n}$ and $\bm{A}_{T_{n}}$ indicates a submatrix of  $\bm{A}$. By allowing the search direction change after each low complexity iteration, CGIHT is able to quickly identify the correct descending direction while using the computational advantages of conjugate gradient method when the support set remains stable.

\begin{algorithm}[h]
\caption{\cite{rj19} Conjugate Gradient Iterative Hard Thresholding restarted for Compressive Sensing}
\label{alg1}
\begin{algorithmic}
\STATE {\textbf{Input:}} ${\bm{A}}$, ${\bm{y}}$, $K$ 
\vspace{1 ex}
\STATE {\textbf{Initialization:}} ${{\bm{c}}_0} = 0$, ${{T}_{-1}} = \emptyset$, 
${{\bm{d}}_{ - 1}} = {0}$, ${T_0} = {\mathop{\rm supp}\nolimits} ({H_K}({{\bm{A}}^T}{\bm{y}}))$.
\vspace{1 ex}
\FOR{each iteration $n \ge 0$}
\STATE 1)\quad${{\bm{g}}_n} = {{\bm{A}}^{\rm T}}({\bm{y}} - {\bm{A}}{{\bm{c}}_n})$\quad (compute the gradient direction) \\
2)\quad \textbf{if}\quad${T_n} \ne {T_{n - 1}}$\quad \textbf{then}\\
\setlength\parindent{3.5em}${\chi _n} = 0$\\
\setlength\parindent{2em}\textbf{else}\\
\setlength\parindent{3.5em}${\chi _n} =  - \frac{{\left\langle {{\bm{A}}{{\bm{g}}_{n{T_n}}},{\bm{A}}{{\bm{d}}_{n - 1{T_n}}}} \right\rangle }}{{\left\langle {{\bm{A}}{{\bm{d}}_{n - 1{T_n}}},{\bm{A}}{{\bm{d}}_{n - 1{T_n}}}} \right\rangle }}$ \quad (compute orthogonalization weight)\\
\setlength\parindent{2em}\textbf{end}\\
\setlength\parindent{0em}3)\quad${{\bm{d}}_n} = {{\bm{g}}_n} + {\chi _n}{{\bm{d}}_{n - 1}}$ \quad (compute conjugate gradient direction)\\
4)\quad${\alpha _n} = {{\left\langle {{{\bm{g}}_{n{T_n}}},{{\bm{g}}_{n{T_n}}}} \right\rangle } \mathord{\left/
 {\vphantom {{\left\langle {{{\bm{g}}_{n{T_n}}},{{\bm{g}}_{n{T_n}}}} \right\rangle } {\left\langle {{{\bm{A}}_{{T_n}}}{{\bm{d}}_{n{T_n}}},{{\bm{A}}_{{T_n}}}{{\bm{d}}_{n{T_n}}}} \right\rangle }}} \right.
 \kern-\nulldelimiterspace} {\left\langle {{{\bm{A}}_{{T_n}}}{{\bm{d}}_{n{T_n}}},{{\bm{A}}_{{T_n}}}{{\bm{d}}_{n{T_n}}}} \right\rangle }}$\quad (compute step size)\\
5)\quad ${{\bm{c}}_{n + 1}} = {H_K}({{\bm{x}}_n} + {\alpha _n}{{\bm{d}}_n})$, ${T_{n + 1}} = {\mathop{\rm supp}\nolimits} ({{\bm{c}}_{n + 1}})$
\vspace{1 ex}
\STATE until the stopping criteria is met
\vspace{1 ex}
\ENDFOR
\label{Algorithm 1}
\end{algorithmic}
\end{algorithm}

Although the CGIHT algorithm can accelerate convergence, the Jacobian matrix needs updating at each iteration in non-linear problems, which may result in search directions that do not satisfy the conjugate condition, resulting in slower convergence. In the process of solving nonlinear radiation flux calculation model, it is time-consuming to update the Jacobian matrix, and we should to devise means to reduce the number of iterations.

\subsubsection{Subspace Pursuit Algorithm}

Orthogonal matching pursuit (OMP) algorithm is wildly used for CS reconstruction owing to its simple implementation and lower computational complexity \cite{rj20}. In each iteration, OMP selects the column of the measurement matrix which is most strongly correlated with the current residuals and adds this column into the set of selected columns, then the residual is updated by projecting the measurement vector $\bm{y}$ onto the linear subspace spanned by the selected columns. The reconstruction complexity of the standard OMP is roughly $O(K M N)$ because it always needs $s$ iterations. Unlike the OMP algorithm, at each iteration, subspace pursuit algorithm selected $s$ candidates based on the correlation values between the columns of $\bm{A}$ and the measurement vector $\bm{y}$, and then a back-tracking operation of the candidate set in the previous iteration is implemented. Compared to the OMP algorithm, the computational complexity of the SP algorithm can be further reduced to $O(M N \log (K))$ when the nonzero entries of the sparse signal decays slowly. The SP algorithm can be described as follows.

\begin{algorithm}[h]
\caption{\cite{rj20} Subspace Pursuit algorithm}
\label{alg2}
\begin{algorithmic}
\STATE {\textbf{Input:}} ${\bm{A}}$, ${\bm{y}}$, $K$, 
\vspace{1 ex}
\STATE {\textbf{Initialization:}} 
${T_0} = \{ K{\rm{\;indices\;of \;the \; largest \; magnitude \; entries \; in \;  the \; vector \;}}{{\bm{A}}^{\rm T}}{\bm{y}}\}$, \\${{\bm{c}}_0} = \arg {\min _{\bm{c}}}{\left\| {{\bm{y}} - {{\bm{A}}_{{T_0}}}{\bm{c}}} \right\|_2}$,
${{\bm{r}}_0} = {\bm{y}} - {{\bm{A}}_{{T_0}}}{{\bm{c}}_0}$.\\
\vspace{1 ex}
\textbf{When} $n \ge 1$ \qquad \textbf{Do}
\vspace{1 ex}
\vspace{1 ex}
\STATE \quad 1)\quad ${\hat T} = T_{n-1} \cup \{{\rm{ indices \; of \; K \; largest \; magnitude \; entries \; of \;  }}{{\bm{A}}^{\rm T}}{\bm{r}_{n-1}}\} $\\ \vspace{1 ex}
\setlength\parindent{1em}2)\quad ${{{\bm{\hat c}}}_n} = \{ \arg \min {\left\| {{\bm{y}} - {\bm{Ax}}} \right\|_2},{\mathop{\rm supp}\nolimits} ({\bm{x}}) \in \hat T\} $ \\ \vspace{1 ex}
\setlength\parindent{1em}3)\quad ${T_n} = {\mathop{\rm supp}\nolimits} ({H_K}({{{\bm{\hat c}}}_n}))$\\ \vspace{1 ex}
\setlength\parindent{1em}4)\quad${{\bm{c}}_n} = \{ \arg \min {\left\| {{\bm{y}} - {\bm{Ax}}} \right\|_2},{\mathop{\rm supp}\nolimits} ({\bm{x}}) \in {T_n}\} $\\ \vspace{1 ex}
\setlength\parindent{1em}5)\quad${{\bm{r}}_n} = {\bm{y}} - {{\bm{A}}_{{T_n}}}{{\bm{c}}_n}$\\
until the stopping criteria is met
\vspace{1 ex}
\vspace{1 ex}
\STATE \textbf{Output:} $\bm{c}_n$, $T_n$
\label{Algorithm 2}
\end{algorithmic}
\end{algorithm}

\subsubsection{Conjugate Gradient Subspace Thresholding Pursuit Algorithm}

Inspired by CGIHT algorithm and SP algorithm, a new greedy iterative algorithm named conjugate gradient subspace thresholding pursuit (CGSTP) is proposed for reconstructing radiation flux. This algorithm is based on SP algorithm, drawing on CGIHT algorithm and using conjugate gradient as the search direction to accelerate convergence. The pseudo-code of CGSTP algorithm is shown in Algorithm \ref{alg3}.

\begin{algorithm}[h]
\caption{Conjugate Gradient Subspace Thresholding Pursuit (CGSTP)}
\label{alg3}
\begin{algorithmic}
\STATE {\textbf{Input:}} ${\bm{A}}$, ${\bm{y}}$, $K$ 
\vspace{1 ex}
\STATE {\textbf{Initialization:}} ${{\bm{c}}_0} = 0$, ${{T}_{-1}} = \emptyset$, 
${{\bm{d}}_{ - 1}} = {0}$, ${T_0} = {\mathop{\rm supp}\nolimits} ({H_K}({{\bm{A}}^T}{\bm{y}}))$.
\vspace{1 ex}
\FOR{each iteration $n \ge 0$}
\STATE 1)\quad${{\bm{g}}_n} = {{\bm{A}}^{\rm T}}({\bm{y}} - {\bm{A}}{{\bm{c}}_n})$\quad (compute the gradient direction) \\
\vspace{1 ex}
2)\quad \textbf{if}\quad${T_n} \ne {T_{n - 1}}$\quad \textbf{then}\\
\setlength\parindent{3.5em}${\chi _n} = 0$\\
\setlength\parindent{2em}\textbf{else}\\
\vspace{1 ex}
\setlength\parindent{3.5em}${\chi  _n} =  - \frac{{\left\langle {{\bm{A}}{{\bm{g}}_{n{T_n}}},{\bm{A}}{{\bm{d}}_{n - 1{T_n}}}} \right\rangle }}{{\left\langle {{\bm{A}}{{\bm{d}}_{n - 1{T_n}}},{\bm{A}}{{\bm{d}}_{n - 1{T_n}}}} \right\rangle }}$ \quad (compute orthogonalization weight)\\
\setlength\parindent{2em}\textbf{end}\\
\setlength\parindent{0em}3)\quad${{\bm{d}}_n} = {{\bm{g}}_n} + {\chi _n}{{\bm{d}}_{n - 1}}$ \quad (compute conjugate gradient direction)\\
\vspace{1 ex}
4)\quad${\mu _n} = {{\left\langle {{{\bm{g}}_{n{T_n}}},{{\bm{g}}_{n{T_n}}}} \right\rangle } \mathord{\left/
 {\vphantom {{\left\langle {{{\bm{g}}_{n{T_n}}},{{\bm{g}}_{n{T_n}}}} \right\rangle } {\left\langle {{{\bm{A}}_{{T_n}}}{{\bm{d}}_{n{T_n}}},{{\bm{A}}_{{T_n}}}{{\bm{d}}_{n{T_n}}}} \right\rangle }}} \right.
 \kern-\nulldelimiterspace} {\left\langle {{{\bm{A}}_{{T_n}}}{{\bm{d}}_{n{T_n}}},{{\bm{A}}_{{T_n}}}{{\bm{d}}_{n{T_n}}}} \right\rangle }}$\quad (compute step size)\\
\vspace{1 ex}
5)\quad $\hat {T}_{n+1} = {\mathop{\rm supp}\nolimits} ({H_K}({{\bm{c}}_n} + {\mu _n}{{\bm{d}}_n})) \cup {T_n}$.\\
\vspace{1 ex}
6)\quad ${{\hat{\bm{c}}}_{n+1}} = \{ \arg \min {\left\| {{\bm{y}} - {\bm{Ax}}} \right\|_2},{\mathop{\rm supp}\nolimits} ({\bm{x}}) \in {\hat{T}_{n+1}}\} $, ${T_{n + 1}} = {\mathop{\rm supp}\nolimits} ({{{\bm{\hat c}}}_{n + 1}})$\\
\vspace{1 ex}
7)\quad ${{\bm{c}}_{n+1}} = \{ \arg \min {\left\| {{\bm{y}} - {\bm{Ax}}} \right\|_2},{\mathop{\rm supp}\nolimits} ({\bm{x}}) \in {T_{n+1}}\} $
\vspace{1 ex}
\STATE until the stopping criteria is met
\vspace{1 ex}
\ENDFOR
\label{Algorithm 3}
\end{algorithmic}
\end{algorithm}

The CGHTP algorithm is a simple combination of the CGIHT algorithm and the SP algorithm, and absorbs the advantages of both algorithms. 
SP algorithm introduces backtracking operation in the iteration process. In each iteration, the support set of the previous iteration is reconsidered, and the new support set is updated by the least square method. For compressible signals, SP algorithm can find the best $s$ atoms in $log(K)$ iteration. However, SP algorithm has strict requirements for RIP conditions.

The CGIHT algorithm optimizes the search direction, and the conjugate direction is used as one of its optimization directions, which has a superlinear convergence rate associated with $(\sqrt{\kappa}-1) /(\sqrt{\kappa}+1)$ \cite{rj25}, where $\kappa$ is the condition number of matrix $\bm{A}$. However, it does not make further accurate estimation of the sparse coefficient after updating the support set, and the CGIHT algorithm only shows its advantage in convergence speed when the  correlation of the columns in $A$ is very small.

The CGIHT algorithm uses the negative gradient direction and the conjugate gradient direction as the search direction. If the sensing matrix remains unchanged in all steps, only a small number of iterations are needed to converge. In fact, since the incomplete orthogonality between the columns of the sensing matrix, the support set selected in each iteration may be different, and the sub-matrix formed by the columns of $\bm{A}$ with the support set as the index is different in each iteration, which will lead to the search directions obtained in the previous iteration and the current iteration are not conjugate, thus the superlinear convergence speed of the algorithm can not be ensured. What's more, in the nonlinear problem, because the CGIHT algorithm does not accurately estimate the sparse coefficients in each iteration, the solution process requires more iterations.

The CGSTP algorithm absorbs the advantages of SP algorithm and CGIHT algorithm in support set update strategy and search direction optimization respectively. CGSTP provides two alternative search directions in the iterative process, including the gradient descent direction and the conjugate gradient direction, and adaptively selects the optimal direction according to the criterion of whether the support set is the same in the previous iteration and the current iteration. In theory, if the sensing matrix in all iterations remains unchanged, CGSTP can achieve superlinear convergence rate \cite{rj25}. However, in solving the non-linear compressed observation model of radiation flux, the Jacobian matrix changes in each iteration, which results in the change of the sensing matrix, and the acceleration effect is not obvious. Fortunately, since CGSTP uses a backtracking operation similar to SP algorithm to accurately estimate the sparse coefficients after each updating of the support set, resulting in a small change in the Jacobian matrix obtained from the previous iteration and the current iteration, which increases the probability that the sensing matrices in different iterations are the same.

\subsection{Overview of radiation flux computation process with CGSTP}

\begin{figure}  
\centering  
\includegraphics[width=.85\textwidth]{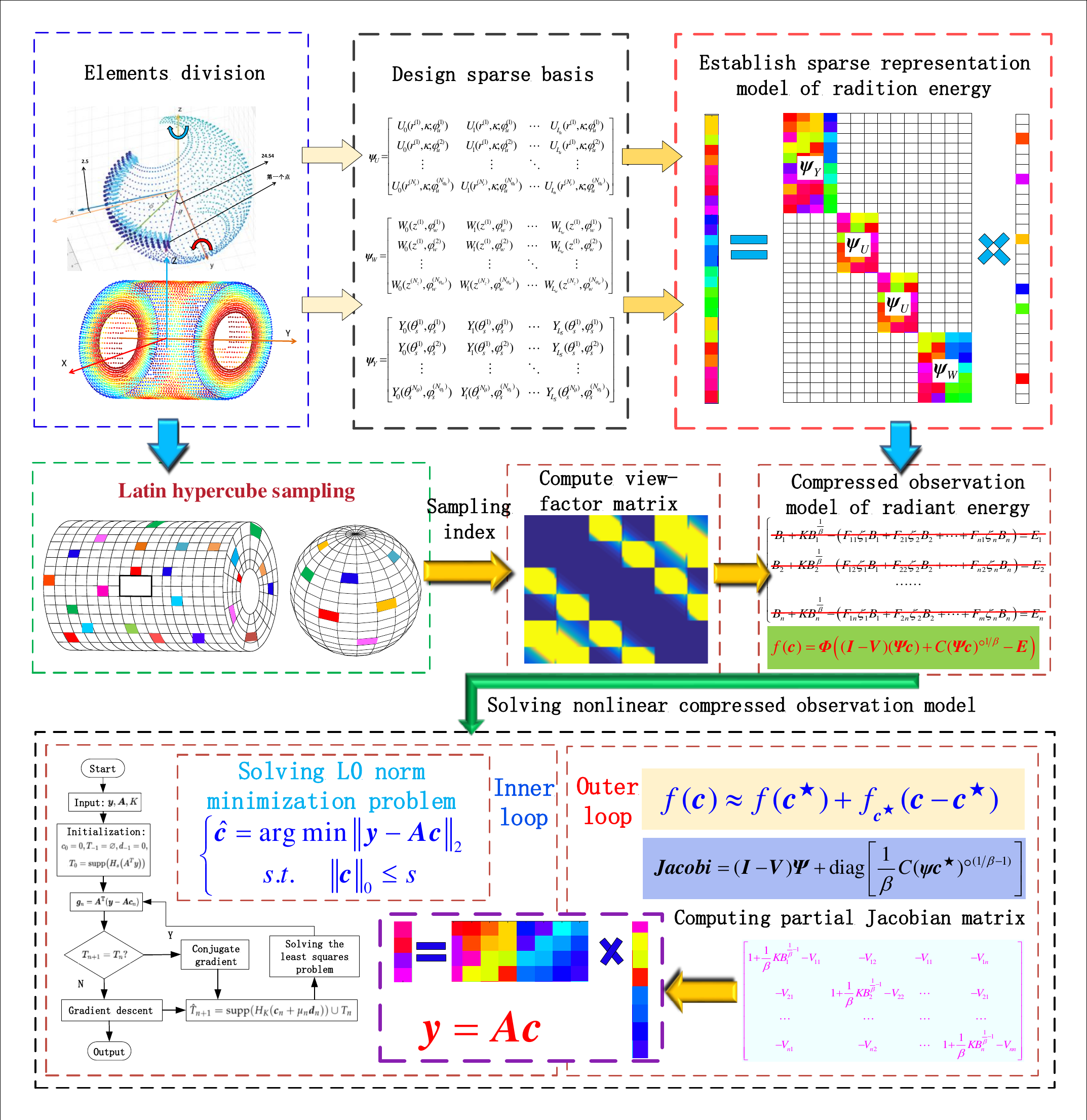}    
\caption{Solution process of nonlinear radiation flux compressed observation model} 
\label{fig6}
\end{figure}

According to the above analysis, solving the non-linear radiation energy balance equation in the Compressed Sensing framework can be summarized into three modules, namely, the sparse representation module of radiation flux, the compressive sampling module and the reconstruction module. Figure \ref{fig6} summarizes the basic contents of each module.

As can be seen from Figure \ref{fig6}, in the sparse representation module, all the radiation regions of the cylindrical cavity model are first divided into discrete elements according to a given precision, and parameter data such as coordinate values, normal vectors, and spatial angles are obtained. Then, the sparse basis composed of orthogonal polynomials of different expansion orders is calculated according to the elements parameters. The representation error of radiation flux over the sparse basis is analyzed to determine the number of expansion terms of polynomials, and the sparsity of radiation flux in the sparse transformation domain is determined by simulation experiments. Finally, a sparse representation model of radiation energy for a small number of sparse coefficients is constructed.

In the compressive sampling module, firstly, all discrete mesh elements are indexed, and $M$ elements are collected by random uniform sampling method (Latin Hypercube Sampling (LHS) method used in this paper). Then, the sparse representation model of radiation flux is transformed into a compressed observation model by using the equation corresponding to $M$ sampling indexes, so as to realize the compression and dimensionality reduction of the model. At the same time, the occlusion condition of each element in the cylindrical cavity is analyzed, and the partial view-factor matrix is calculated according to the sampling index.

In the reconstruction module, the linear approximation system is first constructed by using the first-order Taylor expansion at the current iteration point to construct a linear approximation system. Then, the partial Jacobian matrix is calculated according to the sampling index, and the sensing matrix is calculated according to the Jacobian matrix, The minimization problem based on ${\ell _0}$ norm is established when the sparsity is known as $s$. Finally, the above ${\ell _0}$ norm minimization problem is solved by CGSTP algorithm.

\section{Simulation Experiments and Discussion}

\subsection{Experimental model and simulation environment}

\begin{figure}[htbp]
\centering
\hspace{-.5em}
\subfigure[]
{
\begin{minipage}[t]{0.3\textwidth}
\centering
\includegraphics[width=4cm,height=4cm]{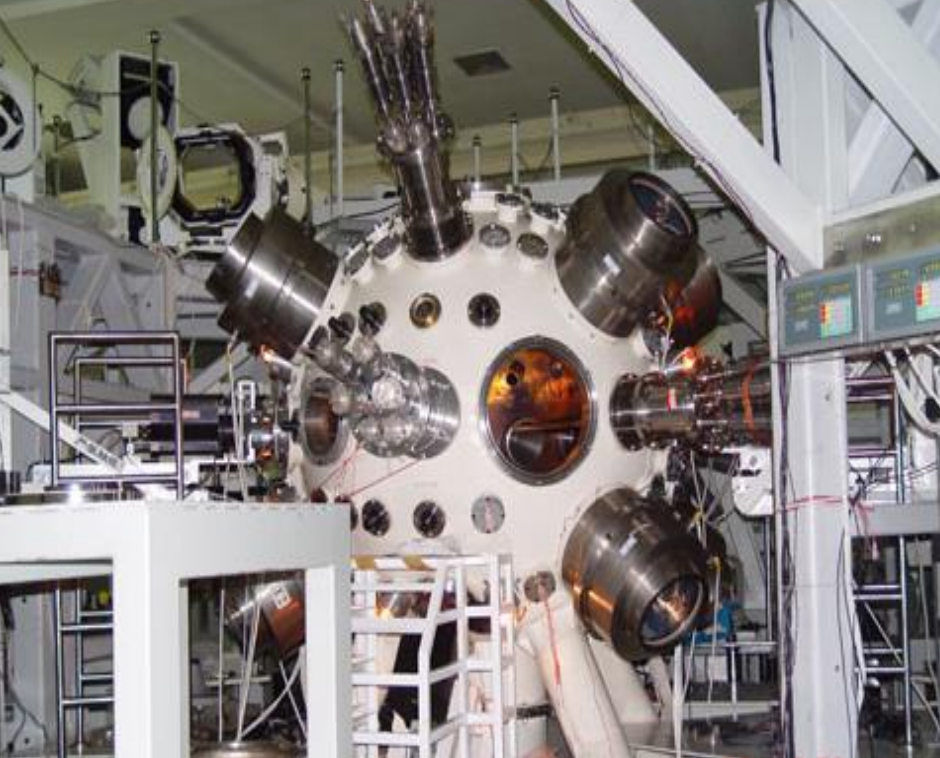}
\end{minipage}
}
\hspace{-2em}
\subfigure[]
{
\begin{minipage}[t]{0.3\textwidth}
\centering
\includegraphics[width=4cm,height=4cm]{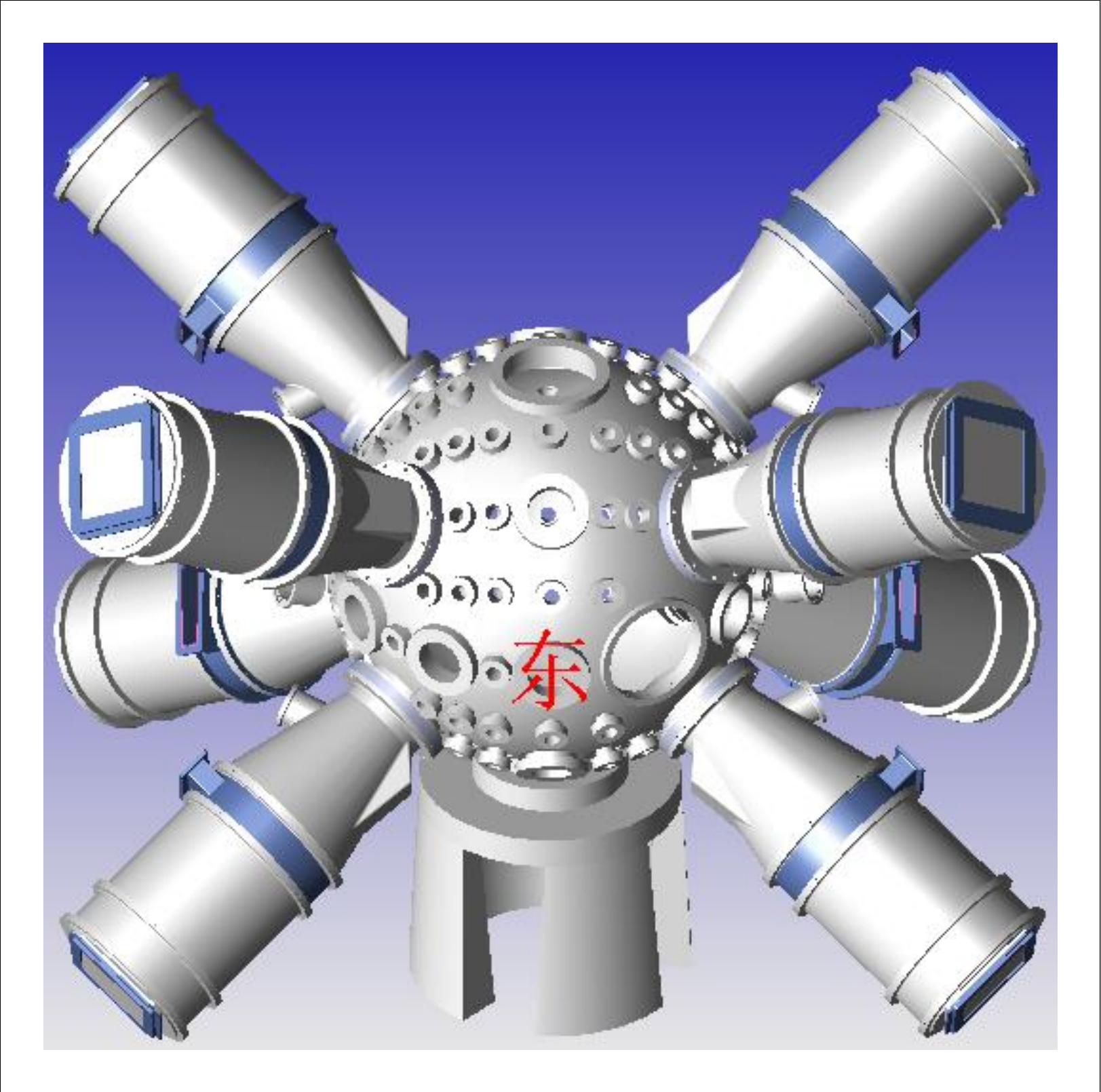}
\end{minipage}
}
\hspace{-2em}
\subfigure[]
{
\begin{minipage}[t]{0.2\textwidth}
\centering
\includegraphics[width=2.1cm,height=4cm]{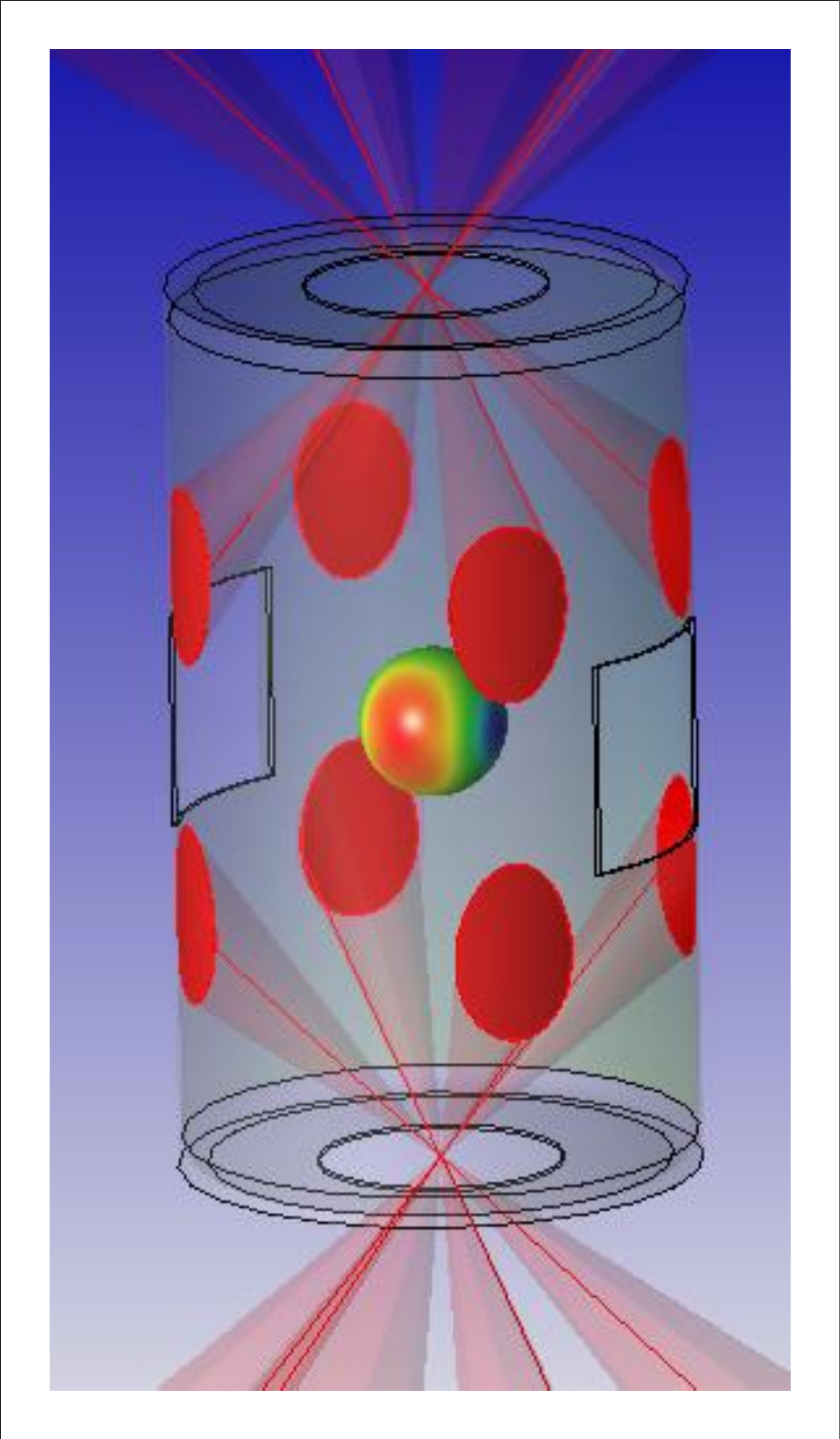}
\end{minipage}
}
\hspace{-2em}
\subfigure[]
{
\begin{minipage}[t]{0.2\textwidth}
\centering
\includegraphics[width=2cm,height=4cm]{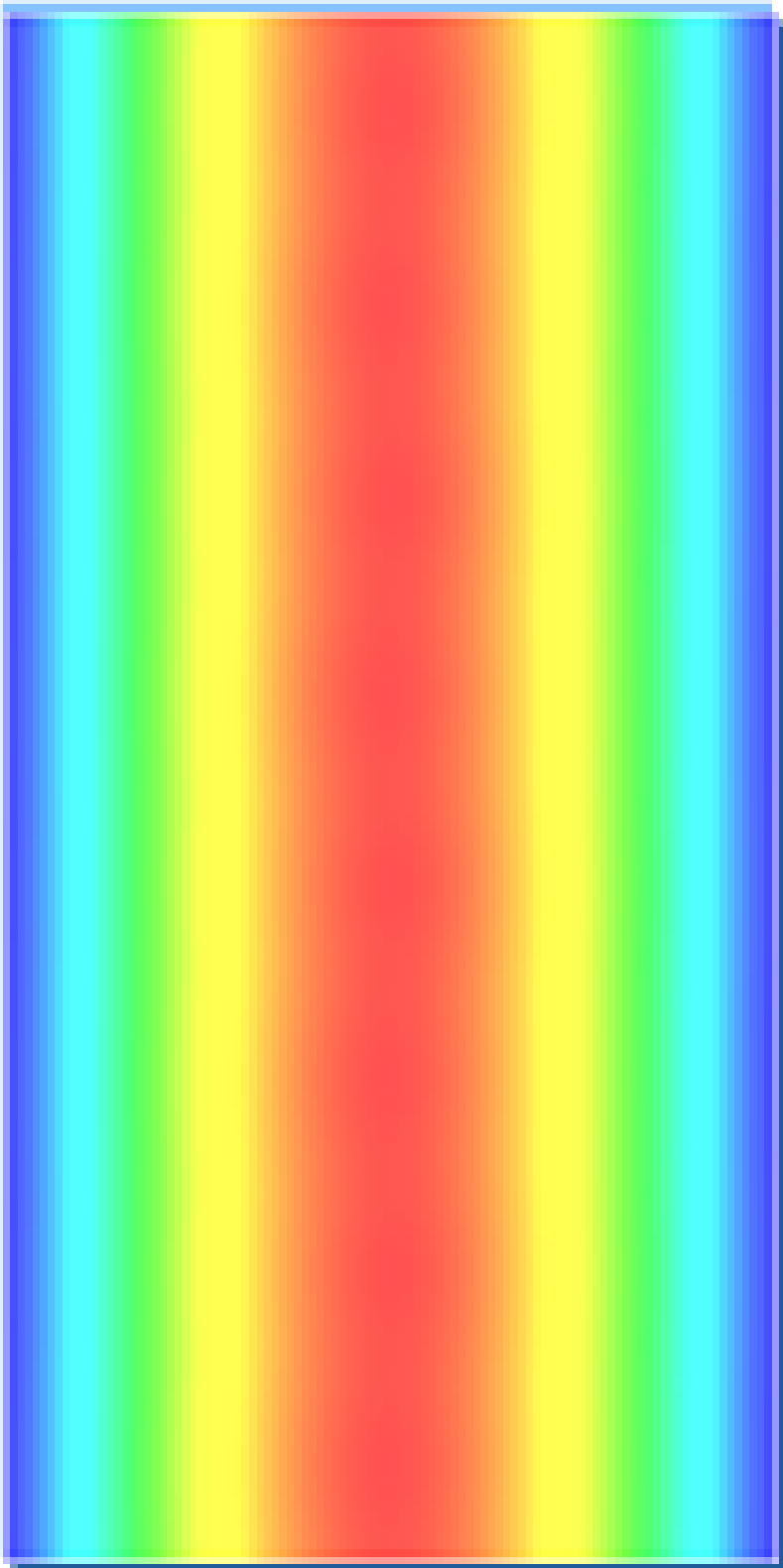}
\end{minipage}
}
\vspace{-1em}
\caption{A cylinder-to-sphere model on the S2 laser facility. \textbf{(a)} S2 laser facility, \textbf{(b)} CAD model of S2, \textbf{(c)} radiation energy computed on the capsule, \textbf{(d)} unfolding view of radiation energy on the capsule}
\label{fig7}
\end{figure}

In this section, two cylindrical cavity models for Shenguang II and Shenguang-III laser devices are used for simulation experiments to verify the performance of the proposed method. The cylindrical cavity model in Shenguang II is shown in Figure \ref{fig7}. The models of Shenguang-II and Shenguang-III are represented by "S2" and "S3" respectively.
All experiments are tested in MATLAB R2015b \cite{rj26}, running on a Windows 10 machine with Intel I5-7500 CPU 3.4GHz and 8GB RAM.

The elements parameters of the S2 model and the S3 model are shown in Table \ref{tab2}, where S2-1, S2-2, S3-1, and S3-2 represent models having different elements sizes and their total elements numbers are respectively 9776, 38952, 20736 and 82944.

The random uniform sampling method used in this section is the Latin Hypercube Sampling (LHS) method. The sampling number is estimated according to the formula $s\log (N)$. Taking the S21 model as an example, the sampling numbers of three radiation regions are calculated as $30 \times \log (2592) \approx 102.4$ (take 150), $35 \times \log (2016) = 115.6$ (take 150) and $100 \times \log (3152) = 349.8$ (take 400).
The sampling rates of different models are shown in Table \ref{tab3}.

As shown in the Table \ref{tab2} and Table \ref{tab3}, we can see that (1) the number of discrete mesh elements significantly increases with the decrease of the size of mesh elements, and (2) the sampling rate for Compressed Sensing based approaches gradually decreases as the number of mesh elements increases.

\newcommand{\tabincell}[2]{\begin{tabular}{@{}#1@{}}#2\end{tabular}}  
\begin{table}[H]
\caption{The size and number of discrete elements of four simulation models. "H", "C", "$\Delta r$" and "$\Delta h$" represent the minimum dimensions of the longitude direction of capsule, the latitudinal direction of capsule, the radial direction of cavity, and the axial direction of cavity, respectively.}
\label{tab2}
\centering
\begin{tabular}{p{10mm}<{\centering}p{30mm}<{\centering}p{35mm}<{\centering}
p{35mm}<{\centering}p{20mm}<{\centering}}
\toprule
\textbf{Model}	& \textbf{Capsule}	& \textbf{The end face of cavity} &\textbf{The cylinder surface} & \textbf{Total number}\\
\midrule
S2-1	 & \tabincell{c}{2592\\(H:$5^\circ$, C:$5^\circ$)}	& \tabincell{c}{2016$\times$2\\(H:$2.5^\circ$, $\Delta r$:15um)}   & \tabincell{c}{3152\\(H:$5^\circ$, $\Delta h$:30um)}          & 9776\\
S2-2	 & \tabincell{c}{10368\\(H:$2.5^\circ$, C:$2.5^\circ$)}			& \tabincell{c}{8064$\times$2\\(H:$1.25^\circ$, $\Delta r$:7.5um)}       & \tabincell{c}{12456\\(H:$2.5^\circ$, $\Delta h$:15um)}          & 38952\\
S3-1   & \tabincell{c}{2592\\(H:$5^\circ$, C:$5^\circ$)}			& \tabincell{c}{4320\\$\times$2(H:$2.5^\circ$, $\Delta r$:15um)}      & \tabincell{c}{9504\\(H:$5^\circ$, $\Delta h$:30um)}          & 20736\\
S3-2	 & \tabincell{c}{10368\\(H:$2.5^\circ$, C:$2.5^\circ$)}			& \tabincell{c}{17280$\times$2\\(H:$1.25^\circ$, $\Delta r$:7.5um)}      & \tabincell{c}{38016\\(H:$2.5^\circ$, $\Delta h$:15um)}         & \tabincell{c}{82944}\\
\bottomrule
\end{tabular}
\end{table}

\begin{table}[H]
\caption{Number of samples of four simulation models}
\label{tab3}
\centering
\begin{tabular}{p{15mm}<{\centering}p{20mm}<{\centering}p{25mm}<{\centering}
p{25mm}<{\centering}p{25mm}<{\centering}p{20mm}<{\centering}}
\toprule
 \multirow{2}{3em}{\textbf{Model}}	& \multirow{2}{3.5em}{\textbf{Capsule}}	& \textbf{The end face of cavity} &\textbf{The cylinder surface} & \textbf{Total sampling number} & \textbf{Sampling rate}\\
\midrule
S2-1	 & 150	 & 300    & 400     & 850  & 8.7\%\\
S2-2  & 150	 & 300    & 450    & 900  & 2.3\%\\
S3-1	 & 150	 & 300    & 350     & 800  & 3.9\%\\
S3-2	 & 150   & 400    & 400    & 950  & 1.1\%\\
\bottomrule
\end{tabular}
\end{table}

\subsection{Sparse representation error analysis of radiation flux}
\begin{figure}[htbp]
\centering
\hspace{-0.5em}
\subfigure[]
{
\begin{minipage}[t]{0.5\textwidth}
\centering
\includegraphics[width=7cm,height=7cm]{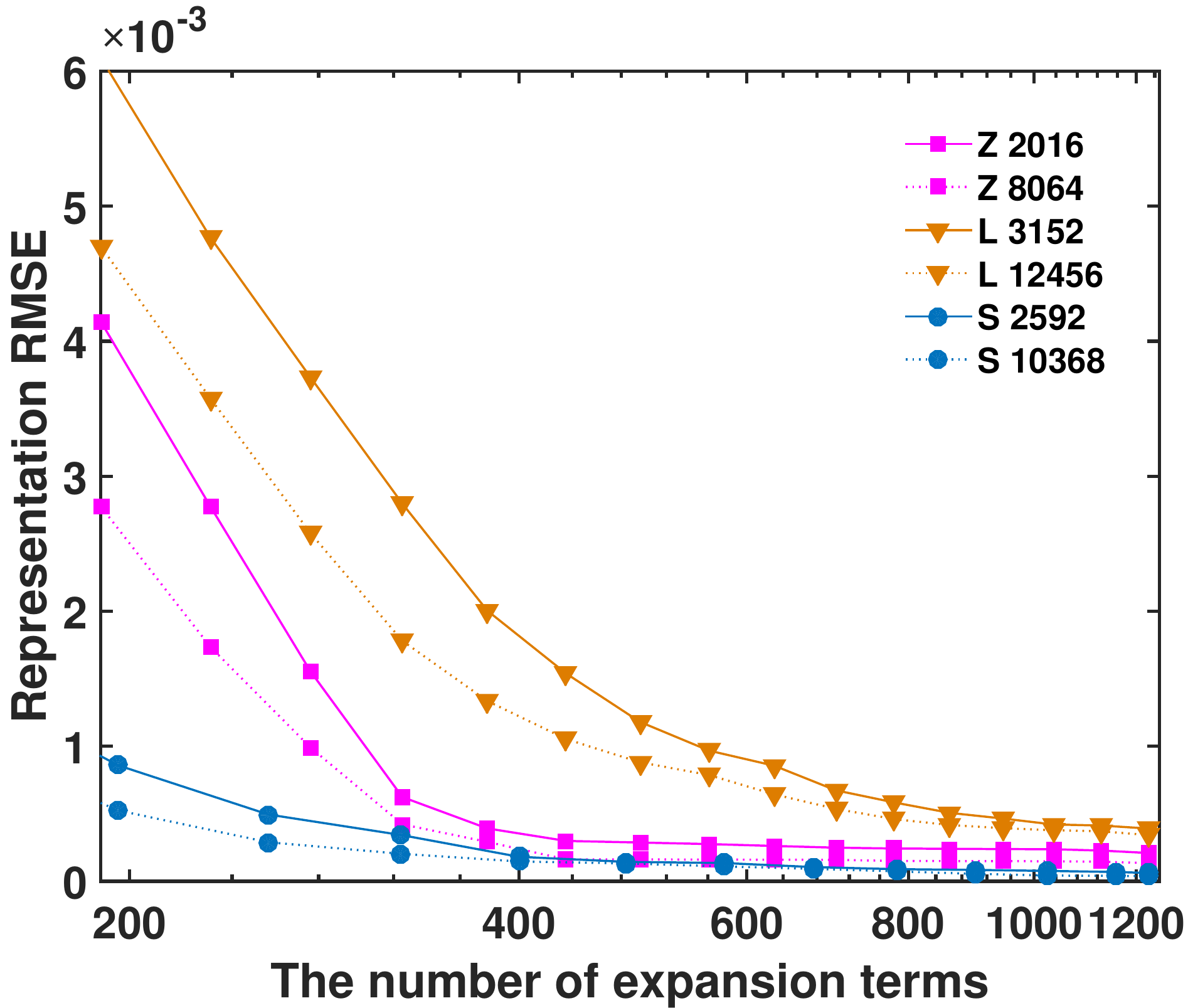}
\end{minipage}
}
\hspace{-2em}
\subfigure[]
{
\begin{minipage}[t]{0.5\textwidth}
\centering
\includegraphics[width=7.3cm,height=7.1cm]{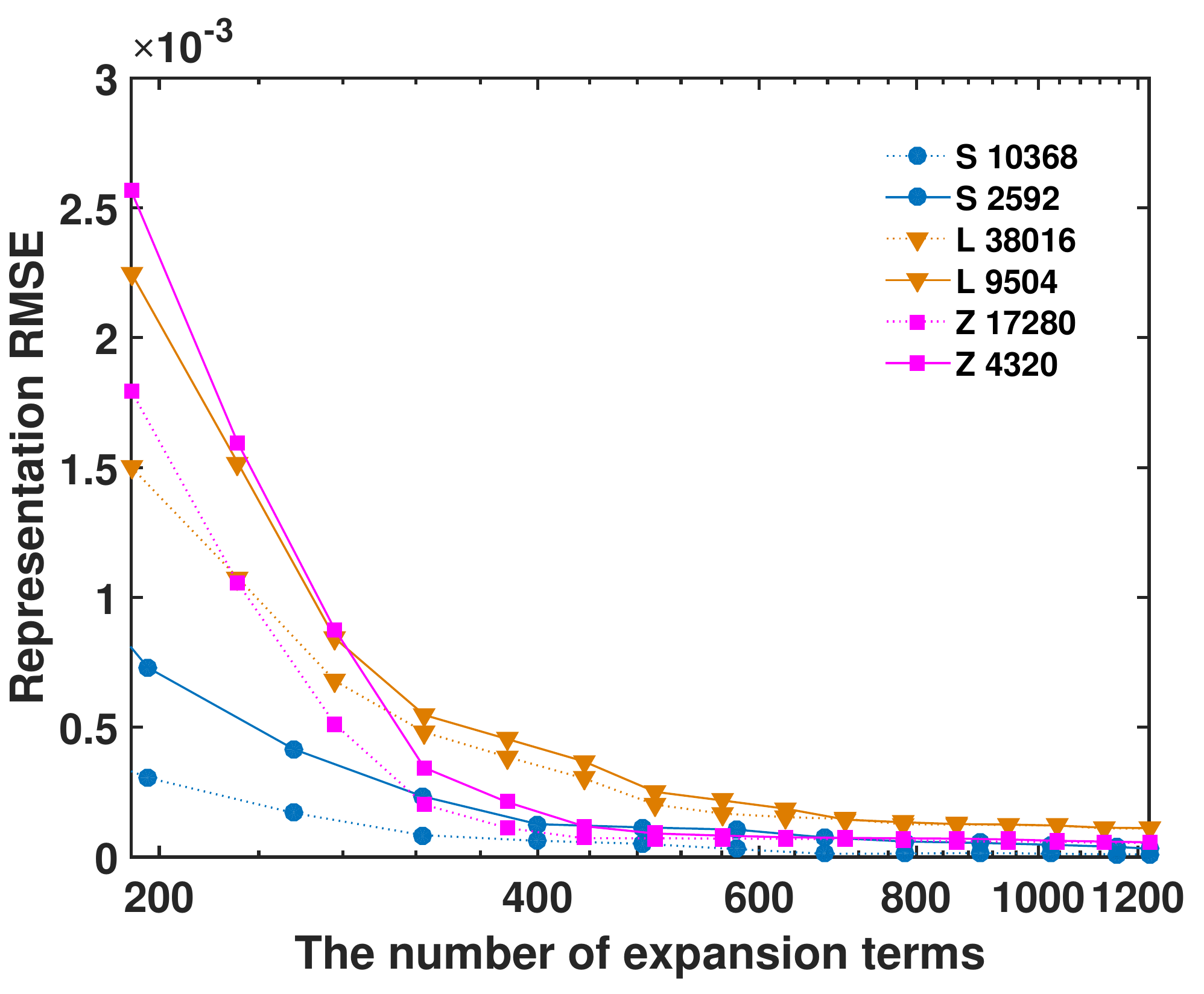}
\end{minipage}
}
\vspace{-1em}
\caption{The relation of representation errors and number of expansions over orthogonal polynomials. "Z 2016", "L 3152" and "S 2592" respectively represent annular Zernike basis with a mesh element number of 2016, Legendre-Fourier basis with a mesh element number of 3152, and a Spherical harmonic basis with a mesh element number of 2016, and the others are similar..  \textbf{(a)} Shenguang-II model, \textbf{(b)} Shenguang-III model.}\label{fig8}
\end{figure}

\begin{figure}[htbp]
\centering
\hspace{-0.5em}
\subfigure[]
{
\begin{minipage}[t]{0.5\textwidth}
\centering
\includegraphics[width=7cm,height=7.1cm]{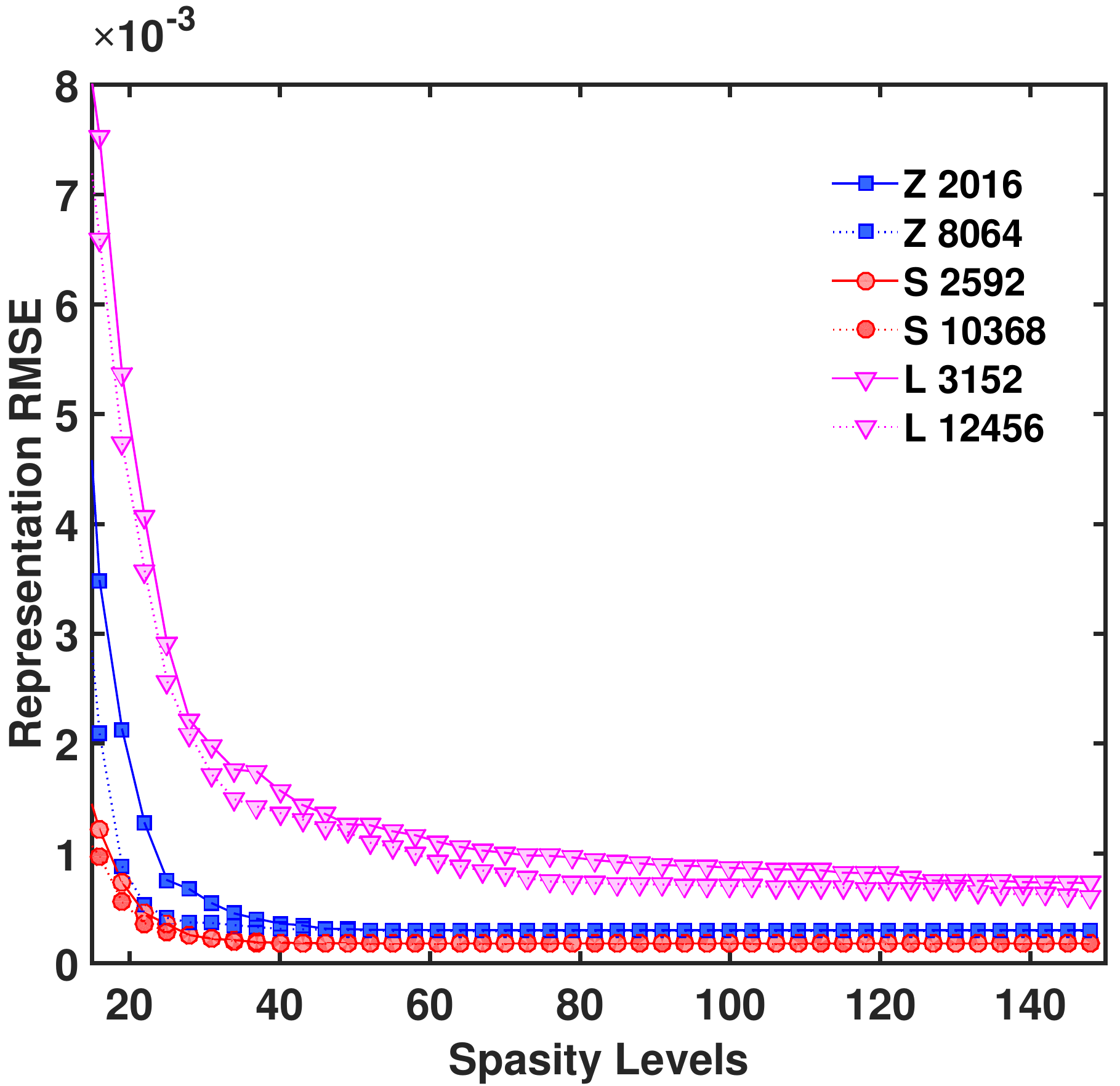}
\end{minipage}
}
\hspace{-2em}
\subfigure[]
{
\begin{minipage}[t]{0.5\textwidth}
\centering
\includegraphics[width=7cm,height=6.8cm]{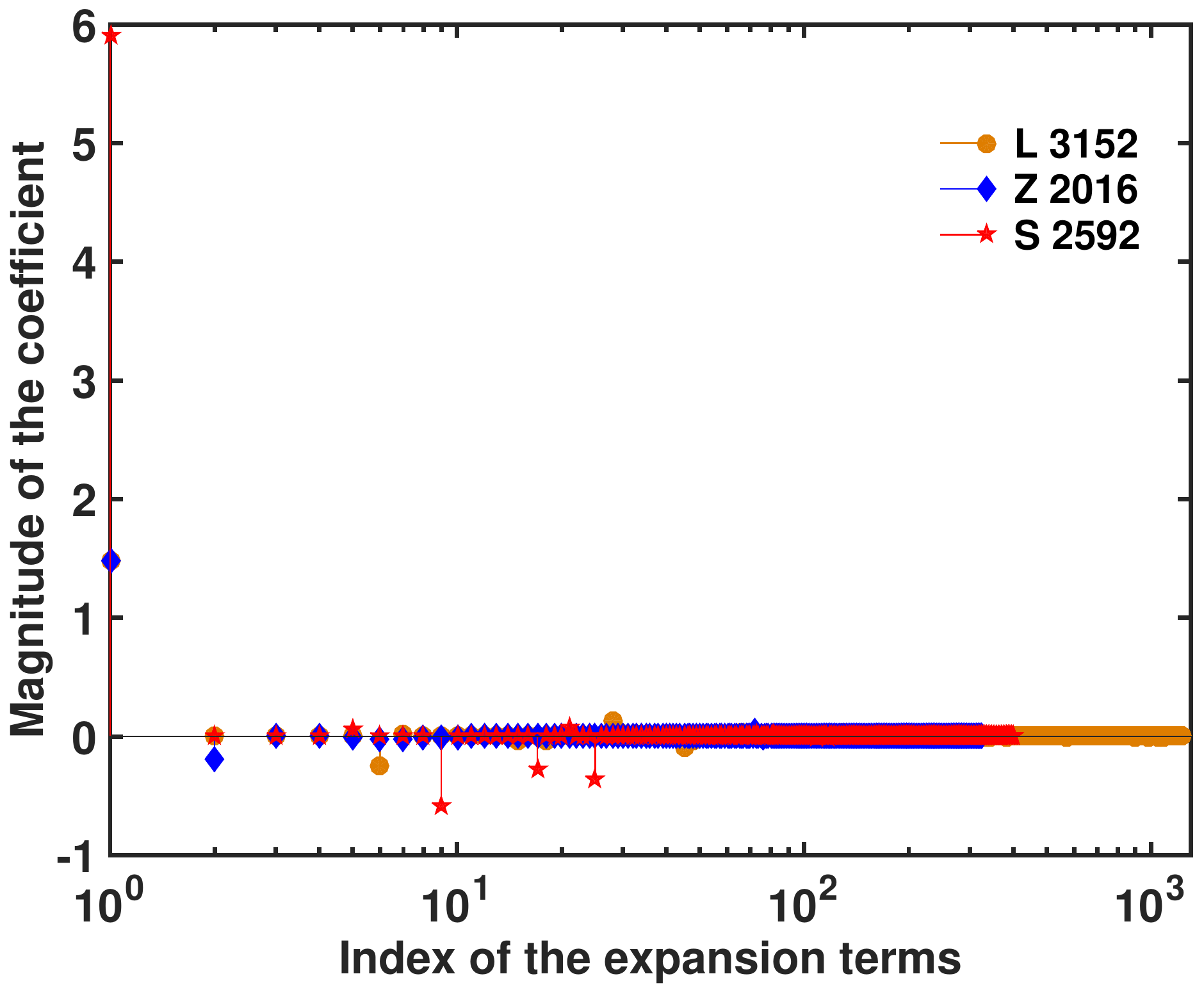}
\end{minipage}
}
\vspace{-1em}
\caption{The sparsity levels over orthogonal polynomials (Shenguang-II model). \textbf{(a)}Representation error v.s. sparsity level, \textbf{(b)}magnitude of the expansion coefficients.}\label{fig9}
\end{figure}

Firstly, we verify the sparsity level of radiation flux ${\bm{B}}_s$, ${\bm{B}}_u$ and ${\bm{B}}_w$ over three orthogonal bases $\bm{Y}$, $\bm{U}$ and $\bm{W}$, respectively.
The expansion coefficients of three sets of polynomials and the representation errors (root mean squared error), between the original flux and the recovered radiation flux with $n_y$, $n_u$ and $n_w$ expansion order are shown in Figure \ref{fig8}(a) and (b) respectively. 
The relationship between the sparsity level of the coefficients and the representation error of the radiant energy flow is shown in Figure 
\ref{fig9}(a) and the magnitude of the sparse coefficients are presented in \ref{fig9}(b).

As shown in Figure \ref{fig8}(a) and (b), with the increase of the number of expansion terms of orthogonal polynomials, the representation error decreases gradually. When the expansion terms of Spherical Harmonic polynomials, Annular Zernike polynomials and Legendre-Fourier polynomials reaches 400, 325 and 1225 respectively, the representation errors approach to zero and the curve remains stable. This means that only a small number of expansion terms are needed to accurately represent the complete radiation flux. From Figure \ref{fig9} (a) we can see that the number of the absolute value of coefficients larger than $10^{-3}$ over three sets of orthogonal bases are $s1 \approx 30$, $s2 \approx 35$ and $s3 \approx 100$ respectively.

Since the top 35 order energy accounts for more than 99.9\% of the total energy and the energy of more than 35th order is little. That is the analysis of radiation flux’s harmonic expansion situation under various irradiation conditions. The drive asymmetry must be less than 2\% and the calculation accuracy should be setting at least an order for magnitude, equal to 0.1\%. 

\subsection{Solution results of radiation energy balance equations}

The resulting radiation flux on the capsule is computed with TDEBM and then compared with Newton-Raphson (NR), Preconditioned Conjugate Gradient (PCG), IHT, CGIHT and CGSTP on computation time, accuracy, iteration number for different kinds of mesh elements sizes. In all tests, the IHT, cgiht, and cgstp algorithms were run independently 20 times, and the average of 20 results was used as the test result.
 Then the computation time list for the four calculation models using different algorithms is shown in Table \ref{tab4} and \ref{tab5}, and the iteration steps and accuracy of different algorithms are shown in Table \ref{tab6} and Table \ref{tab7} respectively.

\begin{figure}[htbp]
\centering
\hspace{-.5em}
\subfigure[]
{
\begin{minipage}[t]{0.5\textwidth}
\centering
\includegraphics[width=7cm,height=6.5cm]{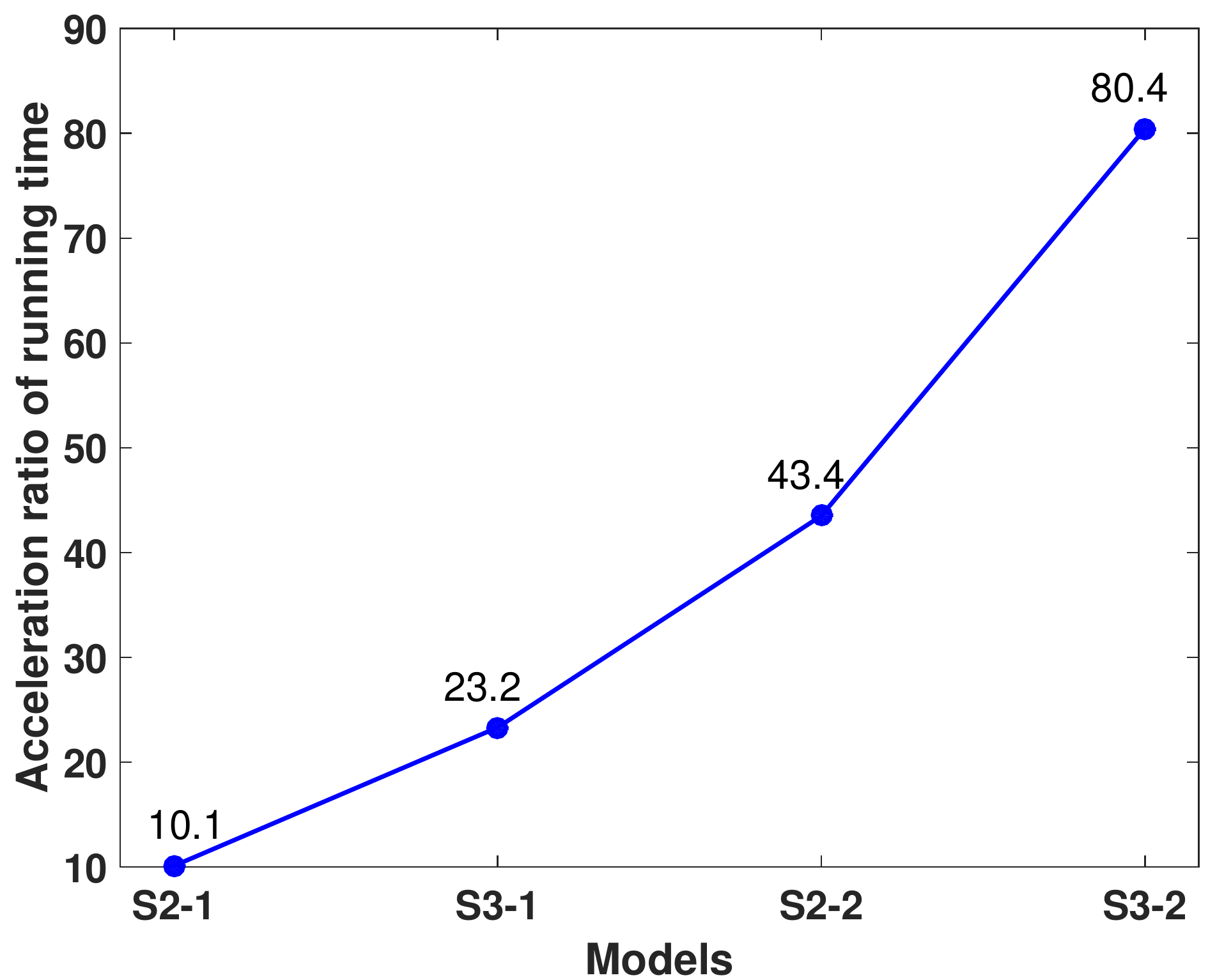}
\end{minipage}
}
\hspace{-2em}
\subfigure[]
{
\begin{minipage}[t]{0.5\textwidth}
\centering
\includegraphics[width=7cm,height=6.5cm]{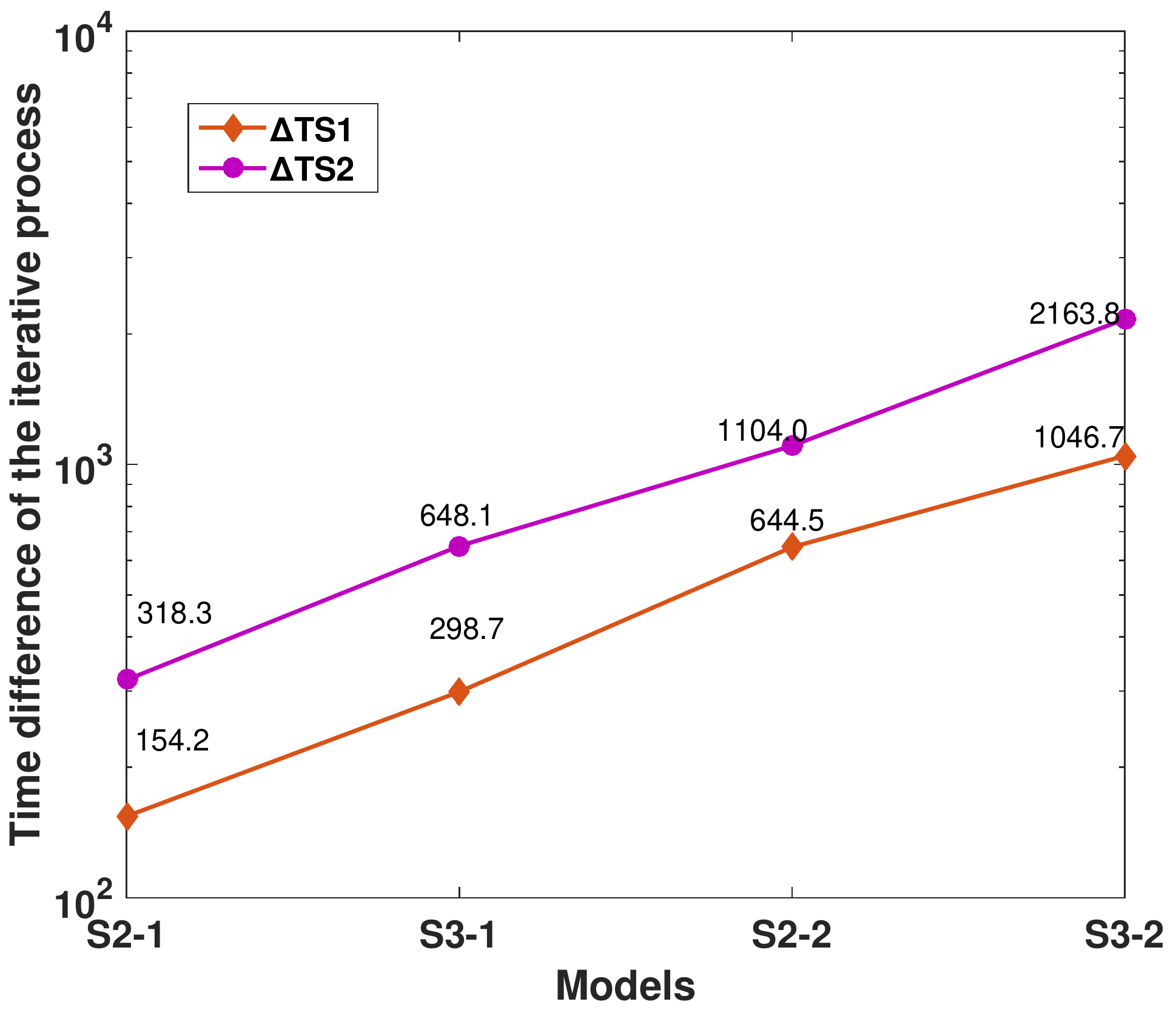}
\end{minipage}
}
\vspace{-1em}
\caption{Running time of different algorithms. \textbf{(a)}The ratio of solving time between CGSTP algorithm and PCG algorithm, \textbf{(b)}$\Delta \rm{TS1}$ and $\Delta \rm{TS2}$ represent the difference of iteration time between CGSTP and CGIHT, CGSTP and IHT algorithm, respectively.}
\label{fig10}
\end{figure}

As shown in Table \ref{tab4} and Table \ref{tab5}, we can see that (1) as the number of non-linear equations exponentially increases, the computation times thus increase sharply for traditional approaches, (2) less computation time is required for compressed sensing based approaches since the computation model is significantly reduced, only almost 8.7\%, or 1.1\% of equations are utilized to obtain the solution, and (3) since the search direction is optimized and the updating strategy of support set is improved, less time required for CGSTP than that of IHT and CGIHT to obtain the solution. Figure \ref{fig10}(a) shows the time ratio between the CGSTP algorithm and the PCG algorithm for calculating the four models. Figure 11(b) shows the iteration time difference between the CGSTP algorithm and the IHT and CGIHT algorithms.
It can be seen from Figure \ref{fig10}(a) that as the model scale increases, the time acceleration ratio of the CGSTP algorithm relative to the PCG method becomes larger and larger, and the maximum acceleration ratio can reach 80. This is because as the scale of the equation increases, the sampling rate becomes smaller and smaller, and the required calculation amount is less and less than the total calculation amount. Figure \ref{fig10} (b) shows that the difference between CGSTP and IHT and CGIHT is larger and larger with the increase of model size. This is because the radiation flux tends to be continuous with the increase of the number of discrete mesh elements, which reduces the number of updates of Jacobian matrix. This means that CGSTP has more advantages when the scale is larger.

\begin{table}[H]
\centering
\caption{Running time of conventional method (unit: second),"--" denotes out of memory}
\label{tab4}
\begin{tabular}{p{20mm}<{\centering}p{30mm}<{\centering}p{20mm}<{\centering}
p{20mm}<{\centering}p{20mm}<{\centering}p{20mm}<{\centering}}
\toprule
  & &\multicolumn{2}{c}{Newton-Raphson method}& \multicolumn{2}{c}{PCG method} \\
\midrule
\multirow{2}{3em}{\textbf{Model}} & \textbf{View-factor calculation time} & \multirow{2}{3em}{\textbf{Iteration time}}  &  \multirow{2}{3em}{\textbf{Total time}} & \multirow{2}{3em}{\textbf{Iteration time}} &  \multirow{2}{3em}{\textbf{Total time}} \\
\hline
S2-1 &   4341.5 &   27.3 &   4368.8 &   21.4  &  4362.9 \\
S2-2 &  69045.8 & 2203.6 &  71249.4 &   313.2 &  69359.1 \\
S3-1 &  16253.1 & 1288.4 &  17541.5 &    96.8 &  16349.9 \\
S3-2 & 255843.2 &    --- &      --- &  1572.6 &  257415.8 \\
\bottomrule
\end{tabular}
\end{table}

\begin{table}[H]
\centering
\caption{Running time of Compressed Sensing method (unit: second).}
\label{tab5}
\begin{tabular}{p{10mm}<{\centering}p{15mm}<{\centering}p{18mm}<{\centering}
p{12mm}<{\centering}p{12mm}<{\centering}p{12mm}<{\centering}
p{12mm}<{\centering}p{12mm}<{\centering}p{12mm}<{\centering}}
\toprule
  & & &\multicolumn{2}{c}{IHT} & \multicolumn{2}{c}{CGIHT}& \multicolumn{2}{c}{CGSTP}\\
\midrule 
\multirow{3}{3em}{\textbf{Model}} & \textbf{View-factor calculation time} & \textbf{Sparse basis calculation time} & \multirow{3}{3em}{\textbf{Iteration time}}  &  \multirow{3}{3em}{\textbf{Total time}} & \multirow{3}{3em}{\textbf{Iteration time}} &  \multirow{3}{3em}{\textbf{Total time}} & \multirow{3}{3em}{\textbf{Iteration time}} &  \multirow{3}{3em}{\textbf{Total time}}\\
\hline
S2-1 &   407.1 &   11.8 &   332.4 &   751.3  &  167.9 &   586.8 & 
13.7      & \textbf{432.6} \\
S2-2 &  1518.4 &   40.2 &   1142.6 &  2701.2 &  683.1 &  2241.7 & 
38.6      & \textbf{1597.2} \\
S3-1 &   658.6 &   22.7 &   671.8 &   1353.1 &  322.4 &  1003.7 & 
20.7      & \textbf{705.0} \\
S3-2 &  3014.2 &  103.5 &  2248.3 &   5366.0 & 1131.2 &  4248.9 &
84.5      &  \textbf{3202.2} \\
\bottomrule
\end{tabular}
\end{table}

\begin{figure}  
\centering  
\includegraphics[width=0.6\textwidth]{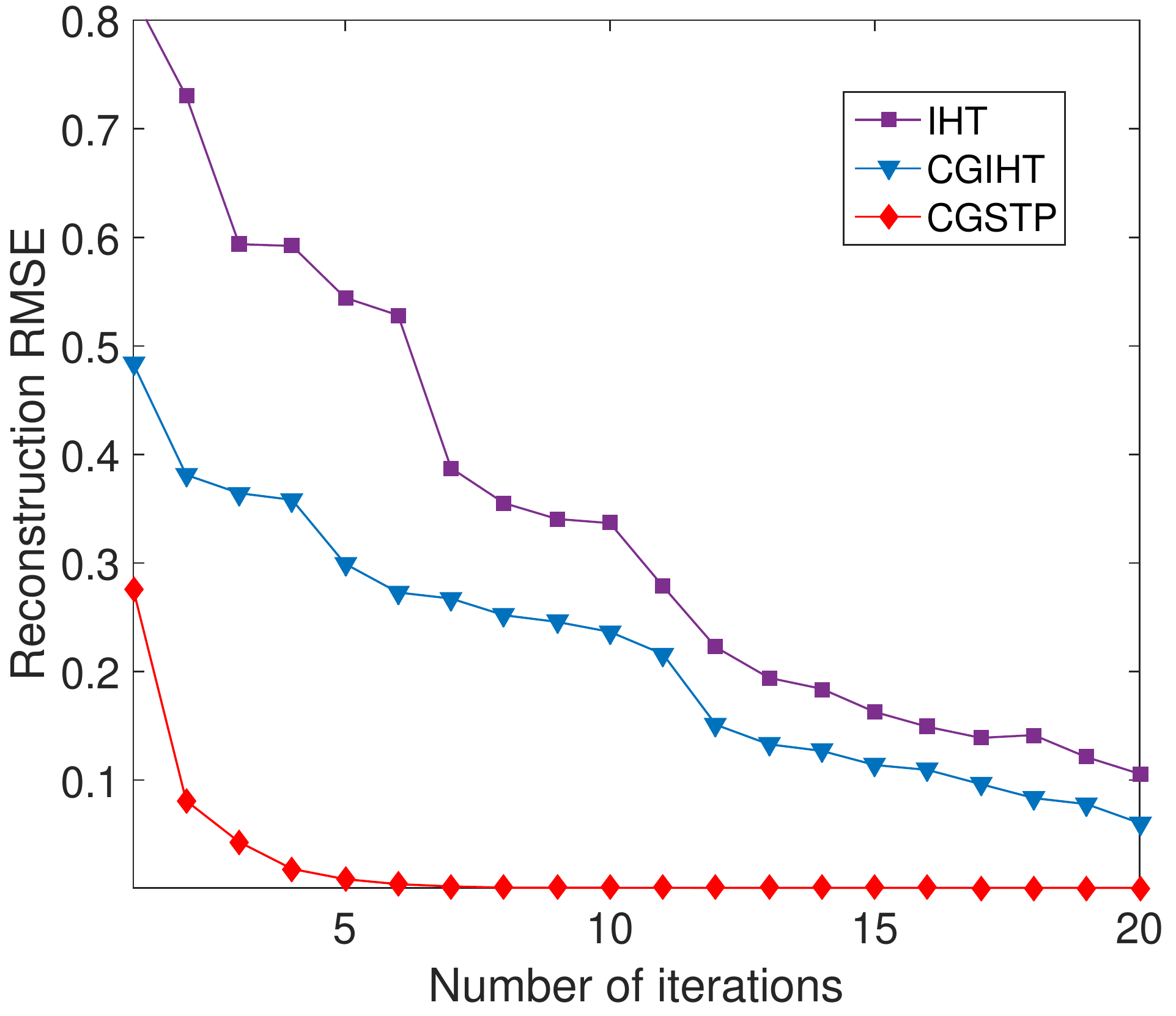}    
\caption{Iteration number and reconstruction error of different algorithms} 
\label{fig11}
\end{figure}

The convergence performance of the above algorithms is shown in Table \ref{tab6}.It can be seen that (1) a small number of iterations is required for the NR, PCG and CGSTP approaches, whereas hundreds of iterations is required for the IHT, CGIHT approaches, (2) Even though less time of iterations are required for the NR and PCG approaches, the computation time for each iteration is much longer than that of IHT, CGIHT and CGSTP algorithms, and (3) as compared with the IHT and CGIHT algorithms, less iterations is required for the CGSTP approach since optimal search direction and step length is applied. Figure \ref{fig11} shows the relationship between IHT, CGIHT and CGSTP algorithm iteration times and reconstruction error. From Figure \ref{fig11}, we can see that the error curve of CGSTP algorithm decreases very fast, and it only needs about six iterations to approach zero and keep stable, which means that the CGSTP algorithms has a faster convergence speed than the other algorithms.

\begin{table}[H]
\caption{Number of samples of four simulation models}
\label{tab6}
\centering
\begin{tabular}{p{20mm}<{\centering}p{25mm}<{\centering}p{20mm}<{\centering}
p{20mm}<{\centering}p{20mm}<{\centering}p{25mm}<{\centering}
}
\toprule
\textbf{Model}	& \textbf{Newton-Raphson}	& \textbf{PCG} &\textbf{IHT} & \textbf{CGIHT} & \textbf{CGSTP}\\
\midrule
S2-1	 & 3	 & 10    & 1273     & 841    & 38\\
S2-2  & 2	 & 8    & 1124    & 729     & 35\\
S3-1	 & 3	 & 9    & 837     & 657    & 31\\
S3-2	 & 2  & 8    & 811    & 594     & 27\\
\bottomrule
\end{tabular}
\end{table}

The accuracy of the different algorithms is listed in Table \ref{tab7}. It can be seen from Table \ref{tab7} that (1) the NR algorithm and the PCG algorithm have the highest accuracy, and the root mean square error (RMSE) is lower than $10^{-5}$, (2) the accuracy of CGSTP algorithm is higher than IHT algorithm and CGIHT algorithm in solving different models, and (3) the accuracy of all algorithms is improved with the increase of the number of mesh elements. The RMSE of CGSTP algorithm in calculating different models is less than $10^{-3}$, which satisfies the radiation symmetry analysis.

\begin{table}[H]
\caption{Reconstruction RMSE of different methods}
\label{tab7}
\centering
\begin{tabular}{p{10mm}<{\centering}p{24mm}<{\centering}p{24mm}<{\centering}p{24mm}<{\centering}
p{24mm}<{\centering}p{24mm}<{\centering}}
\toprule
\textbf{Model}	&\textbf{NR}	& \textbf{PCG} &\textbf{IHT} & \textbf{CGIHT} & \textbf{CGSTP}\\
\midrule
S2-1	 & $3.14\times10^{-10}$ & $5.52\times10^{-6}$ & $3.71\times10^{-3}$ &  $1.53\times10^{-3}$  & $7.84\times10^{-4} $ \\
S2-2  & $2.33\times10^{-10}$ & $3.91\times10^{-6}$ & $1.27\times10^{-3}$ &  $9.84\times10^{-4}$  &  $6.20\times10^{-4}$ \\
S3-1	 & $1.65\times10^{-10}$ & $4.13\times10^{-6}$ & $9.81\times10^{-4}$ &  $8.92\times10^{-4}$  & $5.75\times10^{-4} $ \\
S3-2  & $1.08\times10^{-10}$ & $1.64\times10^{-6}$ & $9.55\times10^{-4}$ &  $8.51\times10^{-4}$  &  $5.24\times10^{-4}$ \\
\bottomrule
\end{tabular}
\end{table}


\section{Conclusions}
The radiation flux distribution symmetry evaluation is very important in ICF experiments, which involve a time consumption process in solving the nonlinear energy equilibrium computation model. In order to accelerate such equation solving process, an efficient radiation computation approach is presented. Firstly, we investigated the distribution characteristics of radiation flux on a cylindrical cavity model and  three sets of orthogonal polynomials are employed to accurately and sparsely represent the radiation flux. Then, only a few mesh elements are sampled to formulate the sparse equation model, which enable the non-linear equation model largely reduced. Finally, a greedy iterative algorithm named CGSTP is presented to efficiently solve the non-linear radiation energy balance equations. Some experimental targets is utilized to validate the efficiency of the presented approach. The result show that the radiation flux can be efficiently obtained in almost ${1 \mathord{\left/
 {\vphantom {1 {80}}} \right.
 \kern-\nulldelimiterspace} {80}}$ time of traditional approach. In future, we will continue to explore the sparse representation for  recent free-form hohlraums, to enable the present approach be used for more targets design, analysis and optimization.

\ifCLASSOPTIONcaptionsoff
  \newpage
\fi

\end{document}